\documentclass[twocolumn,prb,showpacs,aps,superscriptaddress]{revtex4}

\usepackage{amsmath}
\usepackage{amssymb,amsthm}
\usepackage{mathptmx}
\usepackage{graphicx}
\usepackage[usenames,dvipsnames,svgnames,table]{xcolor}
\usepackage[colorlinks=true,linkcolor=blue,
            citecolor=blue,filecolor=blue,baseurl=empty]{hyperref}

\DeclareMathAlphabet{\mathcal}{OMS}{cmsy}{m}{n}

\newcommand{\cds}[1]{\ensuremath{c^\dagger_{#1}}}
\newcommand{\ccs}[1]{\ensuremath{c_{#1}}}

\newcommand{\cd}[1]{\ensuremath{\hat{c}^\dagger_{#1}}}
\newcommand{\cc}[1]{\ensuremath{\hat{c}_{#1}}}

\newcommand{\lwf}[1]{\ensuremath{\langle #1 \vert}}
\newcommand{\rwf}[1]{\ensuremath{\vert #1 \rangle}}
\newcommand{\est}[1]{\ensuremath{\langle {#1} \rangle}}
\newcommand{\dt}[0]{\ensuremath{\mathrm{d}t}}
\renewcommand{\exp}[1]{\ensuremath{\mathrm{exp}\left({#1}\right)}}
\newcommand{\trace}[1]{\ensuremath{\mathrm{tr}\left({#1}\right)}}
\newcommand{\torder}[0]{\ensuremath{\mathcal{T}}}
\newcommand{\tcorder}[0]{\ensuremath{\mathcal{T}_C}}
\newcommand{\deltaloc}[0]{\ensuremath{\delta_\mathrm{loc}}}

\begin{document}

\title{Enforcing conservation laws in nonequilibrium cluster perturbation
theory}

\author{Christian Gramsch} 
\author{Michael Potthoff} 

\affiliation{I. Institute for Theoretical Physics, University of Hamburg,
  Jungiusstra\ss e 9, 20355 Hamburg, Germany}
\affiliation{The Hamburg Centre for Ultrafast Imaging, Luruper Chaussee 149,
  22761 Hamburg, Germany}

\begin{abstract} 
Using the recently introduced time-local formulation of the nonequilibrium
cluster perturbation theory (CPT), we construct a generalization of the
approach such that macroscopic conservation laws are respected.  This is
achieved by exploiting the freedom for the choice of the starting point of the
all-order perturbation theory in the inter-cluster hopping.  The proposed
conserving CPT is a self-consistent propagation scheme which respects the
conservation of energy, particle number and spin, which treats short-range
correlations exactly up to the linear scale of the cluster, and which
represents a mean-field-like approach on length scales beyond the cluster size.
Using Green's functions, conservation laws are formulated as local constraints
on the local spin-dependent particle and the doublon density.  We
consider them as conditional equations to self-consistently fix the
time-dependent intra-cluster one-particle parameters.  Thanks to the intrinsic
causality of the CPT, this can be set up as a step-by-step time propagation
scheme with a computational effort scaling linearly with the maximum
propagation time and exponentially in the cluster size.  As a proof of concept,
we consider the dynamics of the two-dimensional, particle-hole-symmetric
Hubbard model following a weak interaction quench by simply employing two-site
clusters only.  Conservation laws are satisfied by construction.  We
demonstrate that enforcing them has strong impact on the dynamics.  While the
doublon density is strongly oscillating within plain CPT, a monotonic
relaxation is observed within the conserving CPT.
\end{abstract}

\maketitle

\section{Introduction}
\label{sec:intro}

A major challenge of the theory of strongly correlated lattice-fermion models
is to predict the real-time dynamics of local observables on a long time scale.
\cite{PSSV11,ATE+14} Using exact-diagonalization techniques, i.e., full
diagonalization or Krylov-space methods, \cite{HL97} only lattices with a small
number of sites can be addressed such that artificial boundary effects start to
dominate the dynamics after a few elementary hopping processes.  Much larger
systems are in principle accessible by means of quantum Monte-Carlo methods, at
least in thermal equilibrium. \cite{BSS81,WSSB89,AE08} As concerns real-time
dynamics, however, the sign (or complex phase) problem still prevents a
computationally efficient simulation, even for impurity-type models which are
typically sign-problem-free at thermal equilibrium, and despite substantial
progress in the recent past. \cite{MR08,WOM09,SF09,CGRM15} For impurity and for
one-dimensional systems, recent extensions of the numerical renormalization
group \cite{AS05} and of the density-matrix renormalization group
\cite{WF04,Vid04,HLO+16} to the time domain have been shown to be highly
efficient and accurate.

For lattice models in two or higher dimensions, on the other hand, one has to
resort to approximations, e.g., to the time-dependent variational principle
evaluated with Gutzwiller \cite{SSF12} or with Jastrow-like variational wave
functions. \cite{IOI15,CBSF12} Using a Green's-function-based approach, on the
other hand, one may also treat the problem within weak-coupling perturbation
theory.  Naive perturbative techniques, however, usually violate the
macroscopic conservation laws that result from the continuous symmetries of the
lattice-fermion model.  As has been shown by Baym and Kadanoff,
\cite{BK61,Bay62} ``conserving approximations'' can be constructed
diagrammatically by deriving the self-energy from a (truncated) Luttinger-Ward
functional \cite{LW60} involving, e.g., certain infinite re-summations of
diagram classes, and by calculating the single-particle Green's function
self-consistently. Due to the necessary approximation of the $\Phi$ functional,
however, certain low-order diagrams are neglected which implies that, strictly
speaking, such conserving approximations are usually restricted to the
weak-coupling limit. \cite{BSW89,Joura:15}

Nonperturbative conserving approximations can either be constructed with the
help of the many-body wave function,\cite{SSF12,IOI15} or, using Green's
functions, within the framework of the nonequilibrium generalization
\cite{HEAP13} of self-energy-functional theory (SFT). \cite{Pot03a,Pot12} Here,
the Green's function is self-consistently obtained from an optimal self-energy
which makes the grand potential of the initial thermal state, expressed as a
functional of the nonequilibrium self-energy, stationary.  The equilibrium SFT
comprises different approximations, such as the variational cluster
approximation \cite{PAD03,DAH+04} and the dynamical impurity approximation.
\cite{Pot03b} These techniques have been extended to real-time dynamics and
have been applied recently to study the dynamical Mott transition \cite{EKW09}
in the Hubbard model \cite{HEP16a,HEP16b} and a variant of the periodic
Anderson model.  \cite{HP16}

Another nonperturbative conserving approach, which can be derived within the
SFT framework but has actually been proposed much earlier, is the
(nonequilibrium) dynamical mean-field theory. \cite{SM02,FTZ06,ATE+14} Being
the exact theory in the limit of infinite spatial dimensions, conservation laws
are in principle naturally satisfied in this case.  In practice, however, this
requires the exact solution of a highly nontrivial quantum-impurity model out
of equilibrium.  First cluster extensions of the DMFT have been reported as
well. \cite{TBAP14,HTEP16} Those combine the mean-field concept with an
improved description of spatial correlations.

The nonequilibrium extension \cite{BP11,JP13,GP15} of cluster-perturbation
theory (CPT) \cite{SPPL00,GV93} is a strongly simplified variant of a
cluster-embedding approach.  Still, the numerical solution of the basic CPT
equation is complicated by the presence of memory effects which are encoded in
real-time Green's functions within the Keldysh formalism.  This is very similar
to the nonequilibrium Dyson or Kadanoff-Baym equations in other diagrammatic
approaches. As has been shown recently, \cite{GBEK13,BE14} however, the
problem can be mapped exactly onto a noninteracting problem with additional
auxiliary degrees of freedom.  Adopting this idea, we could demonstrate
\cite{GP15} that the CPT real-time dynamics can be understood as a simple
Markovian dynamics of a system of noninteracting fermions but in a much larger
time-dependent bath of virtual degrees of freedom.  Using this reformulation of
the CPT, it has been possible to formally study the real-time dynamics of an
inhomogeneous setup in the two-dimensional Hubbard model consisting of
$10\times 10$ sites up to times of the order of $10^{4}$ where the inverse
nearest-neighbor hopping serves as the time unit. 

Those plain CPT calculations, however, suffer from a couple of conceptual
problems.  The drawback of {\em any} mean-field theory is the missing feedback
of certain correlations on the dynamics of the observables of interest, such
as, e.g., the missing feedback of nonlocal spatial correlations on the local
self-energy in the case of the DMFT.  In the case of plain CPT, the situation
is even worse as there is no feedback at all.  In particular, plain CPT
calculations cannot be expected to respect the macroscopic conservation laws
emerging from the symmetries of the underlying Hamiltonian.  This can be traced
back to the fact that the plain CPT does not contain any element of
self-consistency.  Therefore, it is not surprising that a violation of, e.g.,
total-energy conservation has been observed. \cite{GP15}

With the present study we give a proof of principle that this drawback can be
overcome.  We make use of the fact that the CPT can be viewed as an all-order
perturbation theory \cite{BP11} in the inter-cluster hopping around a system of
decoupled clusters, where the starting point, i.e., the intra-cluster
Hamiltonian, is not at all predetermined.  The idea is to formulate the
macroscopic conservation laws as local constraints on the spin-dependent
particle and doublon density.  These equations are then used to fix the
intra-cluster one-particle parameters and thereby to optimize the starting
point for the cluster-perturbation expansion.  This defines a novel
``conserving cluster-perturbation theory.'' The theory is conserving by
construction, it is nonperturbative, and in principle controlled by the inverse
cluster size as a small parameter.  In practice, however, the accessible
cluster size is limited by the exponential growth of the cluster Hilbert space.
Hence, conserving CPT must be seen as a typical cluster mean-field theory which
correctly accounts for nonlocal correlations up to the linear scale of the
cluster.  
Opposed to standard mean-field theories, the ``mean-field'' or the renormalization of the one-particle parameters is determined by imposing local constraints expressing conservation laws, i.e., it is finally the symmetries of the lattice model which dictates the time-dependent cluster embedding.  
As the theory relies on {\em local} self-consistency or conditional equations, it can easily be extended to inhomogeneous models or inhomogeneous initial states.

While the underlying idea is conceptually simple, its practical realization
requires a couple of new theoretical concepts which are discussed here in
detail.  In particular, the implementation of a causal time-stepping algorithm
requires a careful analysis to which order the renormalization of the
intra-cluster parameters at a certain time slice enters the conditional
equations.  We are able to demonstrate that an efficient numerical
implementation of the theory is possible and discuss first results for weak
interaction quenches in a two-dimensional Hubbard model.  The algorithm scales
linearly with the propagation time and exponentially in the cluster size.
Conservation laws are satisfied with numerical accuracy.  Yet, long time scales
cannot be achieved with the present implementation due to singular points which
are found to evolve during the time propagation. 

The next section briefly states the model and the necessary elements of the
Keldysh formalism.  Section \ref{sec:prep} introduces the CPT and discusses the
formulation of the local constraints.  
The mapping onto a noninteracting auxiliary problem is described in Sec.\ \ref{sec:hamiltonian_based}.  
The main theoretical work addresses the solution of the local constraints for the optimal starting point of the all-order perturbation theory. 
This is presented in Sec.\ \ref{sec:solve_for_lambda}.  
Numerical results are discussed in Sec.\ \ref{sec:num}, and the conclusions are summarized in Sec.\ \ref{sec:con}.

\section{Model and nonequilibrium formalism}
\label{sec:noneq_formalism}

We consider the single-band, fermionic Hubbard model on an arbitrary lattice
with a time-dependent hopping matrix $T(t)$ and interaction strength $U(t)$.
The hopping is assumed as spin-diagonal for simplicity. The Hamiltonian reads
\begin{equation}
  \label{eq:ham}
  H_{T,U}(t) =   \sum_{ij\sigma} (T_{ij\sigma}(t)-\delta_{ij}\mu)
  \cds{i\sigma}\ccs{j\sigma} 
  + U(t)\sum_{i}n_{i\uparrow}n_{i\downarrow},
\end{equation}
where the operators $\cds{i\sigma}$ ($\ccs{j\sigma}$) create (annihilate) a
fermion with spin $\sigma\in \{ \uparrow, \downarrow \}$ at site $i$ ($j$), and
where $n_{i\sigma} = \cds{i\sigma}\ccs{i\sigma}$ denotes the spin-dependent
local density operator. At time $t=0$, the system is assumed to be in thermal
equilibrium with inverse temperature $\beta$ and chemical potential $\mu$.
Nonequilibrium real-time dynamics for $t>0$ is initiated by the time dependence
of the hopping matrix or the interaction strength. In principle, this covers
challenging experimental setups such as time-resolved photoemission
spectroscopy\cite{perf:06} or experiments with ultracold gases in optical
lattices.\cite{bloch:08}

Our central quantity of interest is the one-particle Green's function which
is defined as
\begin{align}
	\label{eq:gf}
  [G_{T,U}]_{ij\sigma}(t,t') 
  &=   
  -i\est{\tcorder\,\cc{i\sigma}(t)\cd{j\sigma}(t')}_{H_{T, U}} \\
  &\equiv
  \frac{-i}{Z}
  \trace{
    \exp{
    -\beta H_{T,U}(0)
    }
    \left[
      \tcorder\,\cc{i\sigma}(t)\cd{j\sigma}(t')
    \right]
  }.\nonumber
\end{align}
Here ``$\text{tr}(\dots)$'' traces over the Fock space, i.e., we take averages
using the grand-canonical ensemble. \makebox{$Z=\trace{\exp{-\beta 
H_{T,U}(0)}}$} defines the grand-canonical partition function and
$\tcorder$ the time-ordering operator on the Keldysh-Matsubara contour $C$. 
The
time variables $t$ and $t'$ are thus understood as contour times. An in-depth
introduction to the Keldysh formalism \cite{kel:64} can be found in Refs.\
\onlinecite{ram:07}, \onlinecite{lee:06}. Throughout the text we use the
convention that operators with a hat carry a time dependence according to the
Heisenberg picture, i.e.,\ $\cc{i}(t) = \mathcal{U}^\dagger(t,0) \ccs{i}
\mathcal{U}(t,0)$, where $\mathcal{U}(t,t') = \torder \exp{-i\int^t_{t'}
H_{T,U}(t_1)\dt_1}$, for $t > t'$, is the system's time-evolution operator and
$\torder$ the time-ordering operator. The dependence of the Green's function on
$T(t)$ and $U(t)$ is made explicit in the notation using subscripts, i.e.,
$G_{T,U}$, where convenient.  A similar notation is also used for other
quantities.

Through Dyson's equation, the Green's function is linked to the self-energy
\begin{equation}
  \label{eq:dyson}
  G_{T,U} = G_{T,0} + G_{T,0} \circ \Sigma_{T,U} \circ G_{T,U},
\end{equation}
where $G_{T,0}$ is the noninteracting propagator. 
Its (contour) inverse is
\begin{equation}        
  [G_{T,0}^{-1}]_{ij\sigma}(t,t') 
               = \left[
                 \delta_{ij}(i\partial_t + \mu) - T_{ij\sigma}(t))
                 \right]\delta_C(t,t').
\end{equation}
In Eq.\ (\ref{eq:dyson}) we made use of the shorthand notation ``$\circ$'' for
the convolution of contour matrices.  In particular 
\begin{equation}
  \label{eq:convolution}
  [\Sigma_{T,U} \circ G_{T,U}]_{ij\sigma}(t,t')
  =
  \int_C \dt_1 
  \sum_{l}
  [\Sigma_{T,U}]_{il\sigma}(t, t_1)[G_{T,U}]_{lj\sigma}(t_1, t') .
\end{equation}
Note that in this context we implicitly assume a contour Dirac delta function
$\delta_C(t,t')$ present in case of time-local quantities.  For example, $T(t)$
should be replaced by $T(t)\delta_C(t,t')$ in a contour convolution, so that
\begin{align}
  [T \circ G_{T,U}]_{ij\sigma}(t,t')
  &=
  \int_C \dt_1 
  \sum_{l}
  T_{il\sigma}(t)\delta_C(t,t_1)[G_{T,U}]_{lj\sigma}(t_1, t')
  \nonumber\\
  &=
  \sum_{l}T_{il\sigma}(t)[G_{T,U}]_{lj\sigma}(t, t').
\end{align}

Combining Dyson's equation with the equation of motion for the one-particle
Green's function, we get
\begin{equation}
  \label{eq:g2l_def}
  [\Sigma_{T,U} \circ G_{T,U}]_{ij\sigma}(t,t')
  = -i U(t) [G_{T,U}^{(2l)}]_{ij\sigma}(t,t'),
\end{equation}
where $G^{(2l)}$ is the two-particle Green's function
\begin{equation}
  \label{eq:g2l_exact}
  [G_{T,U}^{(2l)}]_{ij\sigma}(t,t') 
  = 
  \est
  {
    \mathcal{T}_C\,\hat{n}_{i\bar\sigma}(t)\cc{i\sigma}(t)\cd{j\sigma}(t')
  } , 
\end{equation}
and where $\bar\sigma$ indicates a flip of the spin index $\sigma$, i.e.,
$\bar \uparrow = \downarrow$ and vice versa.  Analogously to Eq.\
\eqref{eq:g2l_exact}, we furthermore have 
\begin{align}        
  \label{eq:g2r_def}
  -i [G^{(2r)}_{T,U}]_{ij\sigma}(t,t') U(t')
  =
  [G_{T,U} \circ \Sigma_{T,U}]_{ij\sigma}(t,t'),
\end{align}
where $G^{(2r)}$ is defined as
\begin{equation}
  \label{eq:g2r_exact}
  [G_{T,U}^{(2r)}]_{ij\sigma}(t,t') 
  = 
  \est
  {
    \mathcal{T}_C\,\cc{i\sigma}(t)\hat{n}_{j\bar\sigma}(t')\cd{j\sigma}(t')
  }.
\end{equation}
The local doublon density $d_i(t)$ can be expressed through the two-particle
Green's functions as 
\begin{align}
  \label{eq:d_is_independent}
  d_i(t) 
  &\equiv
  \est{\hat{n}_{i\uparrow}(t)\hat{n}_{i\downarrow}(t)}_{H_{T,U}}\\
  &= 
  -[G^{(2l)}_{T,U}]_{ii\sigma}(t,t^+) 
  = 
  -[G^{(2r)}_{T,U}]_{ii\sigma'}(t,t^+),\nonumber
\end{align}
with $t^+$ being infinitesimally ``later'' than $t$ in the sense of time
ordering on the Keldysh-Matsubara contour.

\section{Preparations}
\label{sec:prep}

To enforce conservation laws within cluster perturbation theory (CPT), we proceed in two steps.  
First, we work out that the CPT approach is not unique and that there are free parameters at one's disposal.  
Second, to fix these parameters, we suggest to employ local constraints expressing the conservation laws that result from continuous symmetries of the Hubbard model.  
We start with a discussion of the main idea of the CPT and of the local constraints on the spin-dependent particle and doublon density.

\subsection{Conventional cluster perturbation theory}
\label{sec:cpt}

\begin{figure}
\includegraphics[width=0.40\textwidth]{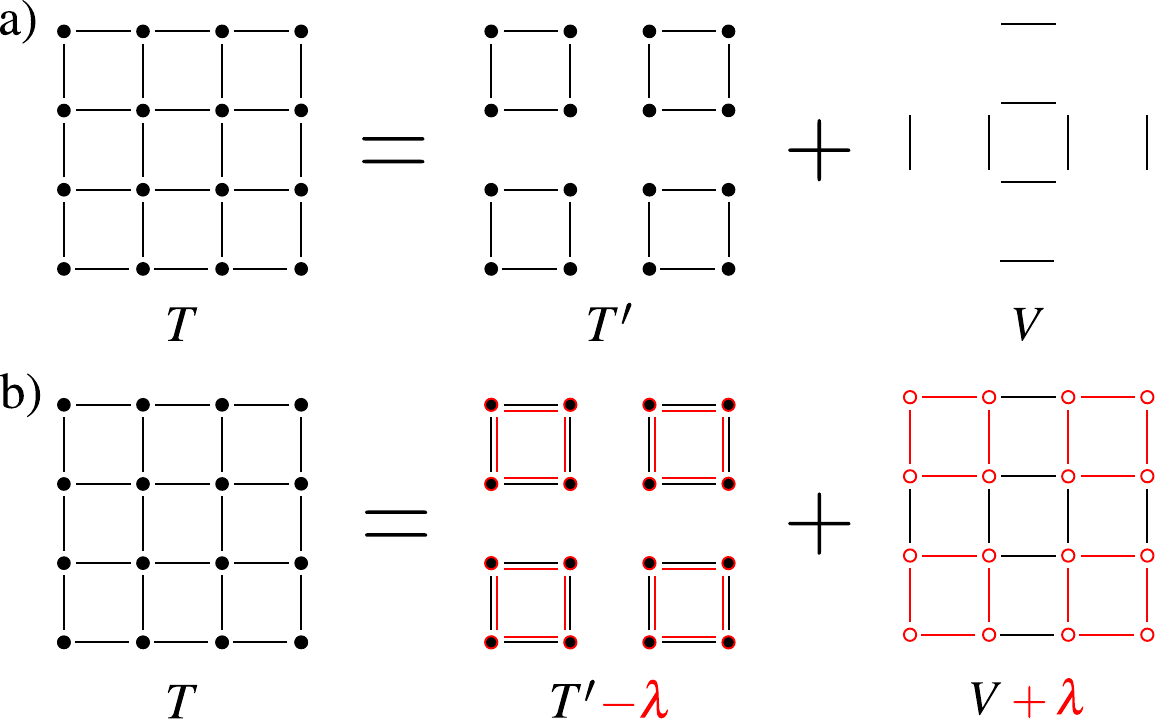}
\caption{
  Sketch of plain CPT (a) and conserving CPT (b).  {\em Plain CPT:} the hopping
  matrix $T$ is decomposed as $T=T'+V$ into the intra-cluster ($T'$) and the
  inter-cluster hopping $V$.  The problem defined by $T'$ (and the local
  Hubbard-type interaction) is solved exactly.  $V$ is treated by all-order
  perturbation theory (neglecting vertex corrections), see Eq.\
  (\ref{eq:cptdyson}).  {\em Conserving CPT:} the same as plain CPT but with
  ``renormalized'' intra- ($T'-\lambda$) and inter-cluster hopping $V+\lambda$,
  where the time-dependent renormalization $\lambda$ (indicated in red) is used
  to enforce conservation laws.  Note that $\lambda$ may also comprise the
  on-site energies.  
}
\label{fig:cpt_vs_optcpt}
\end{figure}

The idea of the CPT \cite{SPPL00,GV93} is to partition the lattice into
clusters small enough to be treated exactly, e.g., using Krylov-space methods
or full diagonalization, and to subsequently include the connections between
the clusters perturbatively. On the level of the Hamiltonian one starts by
partitioning the full hopping matrix $T$ into the intra-cluster hopping $T'$
and the inter-cluster hopping $V$ so that $T'$ only contains terms which
connect lattice sites within the individual clusters, while $V$ contains all
remaining terms such that $T=T'+V$, see Fig.\ \ref{fig:cpt_vs_optcpt}.
Corresponding to the intra-cluster hopping, we define a cluster Hamiltonian
$H_{T', U}(t)$ which describes the system of isolated clusters, also referred
to as the reference system.  Its Green's function and self-energy are denoted
as $G_{T', U}$ and $\Sigma_{T', U}$, respectively.  

For the equilibrium as well as for the nonequilibrium case,
\cite{BP11,JP13,GP15} the CPT can be seen as an all-order perturbation theory
in the inter-cluster hopping $V$ which provides the one-particle Green's
function of the original system by expanding around the cluster Green's
function:
\begin{align}
	\label{eq:cptdyson}
  G^\mathrm{CPT} 
  = 
  G_{T', U} + G_{T', U}\circ V\circ G_{T', U} + \dots
  = 
  \frac{1}{G_{T', U}^{-1}-V} . 
\end{align}
We also have:
\begin{equation}  
  G^\mathrm{CPT} 
  = 
  \frac{1}{G_{T,0}^{-1}-\Sigma_{T', U}} . 
\end{equation}
In the noninteracting case, this is exact since $\Sigma_{T', 0}=0$.  For finite
$U(t)$, however, the CPT Green's function $G^\mathrm{CPT}$ represents an
approximation of the exact Green's function $G_{T,U}$.

A closer look reveals that the CPT is not unique since one may consider a
different starting point for the all-order perturbation theory in $V$. To this
end, consider a starting point with a renormalized intra-cluster hopping, $T'
\rightarrow T' - \lambda$, resulting in a renormalized cluster Green's function
$G_{T' - \lambda, U}$ and self-energy $\Sigma_{T' - \lambda, U}$.
Correspondingly, also the inter-cluster hopping $V$ must be renormalized as $V
\rightarrow V + \lambda$. Summation of the geometrical series yields
\begin{align}
  \label{eq:cptlambda}
  G^\mathrm{CPT}[\lambda]
  = 
  \frac{1}{G_{T'-\lambda, U}^{-1}-(V+\lambda)}
  = 
  \frac{1}{G_{T,0}^{-1}-\Sigma_{T' - \lambda, U}},
\end{align}
where we emphasized the special role of the renormalization parameter $\lambda$
by square brackets. For $U(t) = 0$, we have $G^\text{CPT}[\lambda] =
G_{T,0}$ for any $\lambda$. For an interacting system, however, the choice
for $\lambda$ is crucial, i.e., the resulting CPT Green's function does depend
on the starting point of the all-order perturbation theory in the inter-cluster
hopping. 

This ambiguity in the definition of the CPT seems to be problematic on first
sight, yet it can be turned into an advantage by interpreting the
renormalization $\lambda$ as an optimization parameter.  This has been worked
out systematically in the context of the (nonequilibrium) self-energy
functional theory (SFT), \cite{HEAP13,HEP16a,HEP16b,HP16} where the optimal
$\lambda$ is derived from a variational principle based on the self-energy.
Here, we will take a different route and use the freedom in $\lambda$ to enforce the local constraints on spin-dependent particle and doublon density. 
Physically, the parameter set $\lambda$ must be interpreted as a nonlocal mean-field and the resulting conserving CPT as a cluster mean-field theory.

\subsection{Formulation of the conservation laws as local constraints}
\label{sec:con_eqs}

While conservation laws like particle-number or energy conservation are
naturally fulfilled if one is able to treat a physical problem exactly, this is
not necessarily the case when working with approximate methods. For
Green's-function-based methods it was shown by Baym and Kadanoff
\cite{BK61,Bay62} that respecting certain symmetry relations for the
two-particle Green's function is sufficient to ensure that an approximation is
conserving. 

Here, we build on an equivalent formulation of the macroscopic conservation laws for the particle number, spin and energy and reformulate them as local constraints for the spin-dependent particle density and the doublon density, respectively. 
This is in the spirit of expressing conservation laws of a classical field theory as continuity equations and follows the work of Baym and Kadanoff. \cite{BK61,Bay62} 
One should note, however, that in our case the local constraints cannot be written in the standard form of continuity equations, as here we aim at an approach for a discrete lattice model.

To discuss the local constraints, we first consider the exact time evolution of a system described by the Hubbard Hamiltonian $H_{T,U}(t)$. 
We write $G \equiv G_{T,U}$, $G^{(2l)} \equiv G^{(2l)}_{T,U}$ and $G^{(2r)} \equiv G^{(2r)}_{T,U}$ in this subsection to keep the notation simple.
The exact time evolution of the system will preserve the total particle number and the $z$-component of the total spin as can be expressed by the following local constraint for the spin-dependent density:
\begin{align}
  0
  &=
  \partial_t \est{\hat{n}_{i\sigma}(t)}_{H_{T,U}}
  -
  [G \circ T - T \circ G]_{ii\sigma}(t,t^+),\nonumber\\
  \Leftrightarrow\quad
  F_{i\sigma}(t)
  &\equiv 
  G^{(2l)}_{ii\sigma}(t,t^+) - G^{(2r)}_{ii\sigma}(t,t^+) = 0 \; , 
  \label{eq:con_dens}
\end{align}
as can be verified directly using Eq.\ \eqref{eq:d_is_independent}.

The first line of Eq.\ \eqref{eq:con_dens} constitutes the discrete-lattice analog of the continuity equation for the spin-dependent particle density.
Opposed to a continuum theory, however, the divergence of the spin-dependent particle-current density is replaced by the commutator. 
The second line of Eq.\ \eqref{eq:con_dens} is an equivalent formulation of the same constraint as has originally been mentioned by Baym and Kadanoff. \cite{BK61,Bay62} 

Next, we consider the following local constraint for the doublon density [cf.\ Eq.\ \eqref{eq:d_is_independent}]:
\begin{align}
  \label{eq:con_energy}
  C_{i\sigma}(t)
  \equiv \,\,
  &i\partial_t
  \left[
    G^{(2l)}_{ii\sigma}(t,t^+)
    +
    G^{(2r)}_{ii\sigma}(t,t^+)
  \right]\\
  &
  -
  2\sum_{j\sigma}
  \left[
    T_{ij\sigma}(t)G^{(2r)}_{ji\sigma}(t,t^+)
    -
    G^{(2l)}_{ij\sigma}(t,t^+)T_{ji\sigma}(t)
  \right] = 0.\nonumber
\end{align}
In the exact theory, this constraint together with the above constraint $F_{i\sigma}(t)=0$ expresses the necessity that the doublon density can be derived consistently from either $G^{(2l)}$ or $G^{(2r)}$ and for each spin component $\sigma$ in Eq.\ \eqref{eq:d_is_independent}. 

More important, in case of a time-independent Hamiltonian, i.e., if $H_{T,U}(t)=\mathrm{const.}$ for $t > t_0$, Eq.\  (\ref{eq:con_energy}) implies total-energy conservation. 
This is explicitly shown in the Appendix \ref{ap:con_double_occ} where, for completeness, also a formal derivation of
Eq.\ \eqref{eq:con_energy} is carried out.

While in the exact theory the equations $F_{i\sigma}(t)=0$ and $C_{i\sigma}(t)=0$ must hold necessarily, this is no longer guaranteed in an approximate approach. 
In particular, the equations are usually violated within the conventional CPT. 

The important point is that via Eqs.\ (\ref{eq:g2l_def}) and (\ref{eq:g2r_def}) both, $G^{(2l)}$ and $G^{(2r)}$, can be expressed in terms of the single-particle Green's function and the self-energy and thus both equations $F_{i\sigma}(t)=0$ and $C_{i\sigma}(t)=0$ can be expressed in terms of the central quantities of the CPT. 
Furthermore, as is shown below, they can be incorporated in the Markovian time-propagation scheme based on the Hamiltonian formulation of the CPT. 
The latter is essential for the numerical treatment.

Our main idea is thus to enforce the local constraints $F_{i\sigma}(t)=0$ and $C_{i\sigma}(t)=0$ within the context of the CPT by exploiting the above-discussed freedom in the choice of the CPT starting point, i.e., by choosing an appropriate renormalization $\lambda = \lambda^\mathrm{opt}$.
If $\lambda^\mathrm{opt}$ can be found, this automatically ensures the conservation of particle number, spin and energy. 

\section{Hamiltonian-based formulation}
\label{sec:hamiltonian_based}

In the last section we have introduced the CPT in its usual form, i.e., based
on the self-energy $\Sigma_{T'-\lambda, U}$ of the reference system and Dyson's
equation. A major drawback of this approach is its limitation for the maximum
propagation time that can be reached in a practical numerical calculation. This
is due to the fact that the CPT Green's function and the self-energy of the
reference system are nonlocal in time through their dependence on two contour
times.  The necessary storage for these quantities scales quadratically with
the maximum propagation time, the effort for solving Dyson's equation scales
cubically.  This intrinsic limitation can be overcome if a so-called
Lehmann representation of the self-energy is available. This allows
us to solve the Dyson equation by means of a Markovian propagation scheme which
permits to reach much longer time scales. \cite{BE14,GP15} In the
following, we consider this Lehmann representation of the self-energy as given.
Its existence for an arbitrary, fermionic lattice system out of equilibrium has
been shown in Ref.\ \onlinecite{GP15} where also a constructive numerical
scheme has been presented. It can be used in case of small clusters
accessible to exact-diagonalization techniques. In the following we briefly
recall the main results and then discuss the application to conservation laws
and the respective local constraints.

\subsection{Convolution-free definition of $G^{(2l)}$ and $G^{(2r)}$}

The nonequilibrium self-energy $\Sigma \equiv \Sigma_{T, U}$ of any
lattice-fermion model has a unique Lehmann representation: \cite{GP15}
\begin{align}
  \label{sig:leh}
  \Sigma_{ij\sigma}(t,t')
  &=
  \delta_C(t,t') \Sigma^\mathrm{HF}_{ij\sigma}(t)
  + 
  \sum_{s\sigma} h_{is\sigma}(t) g(h_{ss\sigma};t,t') h^*_{js\sigma}(t') \: .
\end{align}
Here, $\Sigma_{ij\sigma}^\mathrm{HF}(t)$ is the time-local Hartree-Fock term.
The second term has a hybridization-function-like structure \cite{GBEK13,
BE14} where $h_{is\sigma}(t)$ denotes the hopping between a physical site
$i$ and an additional \emph{virtual} site labeled by the index $s$. The
time-independent on-site energy of the virtual site is given by $h_{ss}$.
Furthermore, $g(\epsilon;t,t')$ is the noninteracting Green's function of an
isolated one-particle mode ($h_\mathrm{mode} = \epsilon \cds{}\ccs{}$) with
excitation energy $\epsilon$:
\begin{equation}
  \label{eq:isolated}
  g(\epsilon;t,t')=i[f(\epsilon)-\Theta_C(t,t')] e^{-i\epsilon(t-t')}.
\end{equation}
Here, $f(\epsilon)=(e^{\beta \epsilon}+1)^{-1}$ denotes the Fermi-function,
and $\Theta_C(t,t')$ refers to the contour variant of the Heaviside step
function, i.e., $\Theta_C(t,t')=1$ for $t\ge_Ct'$, and $\Theta_C(t,t')=0$
otherwise.

\begin{figure}
  \includegraphics[width=0.45\textwidth]{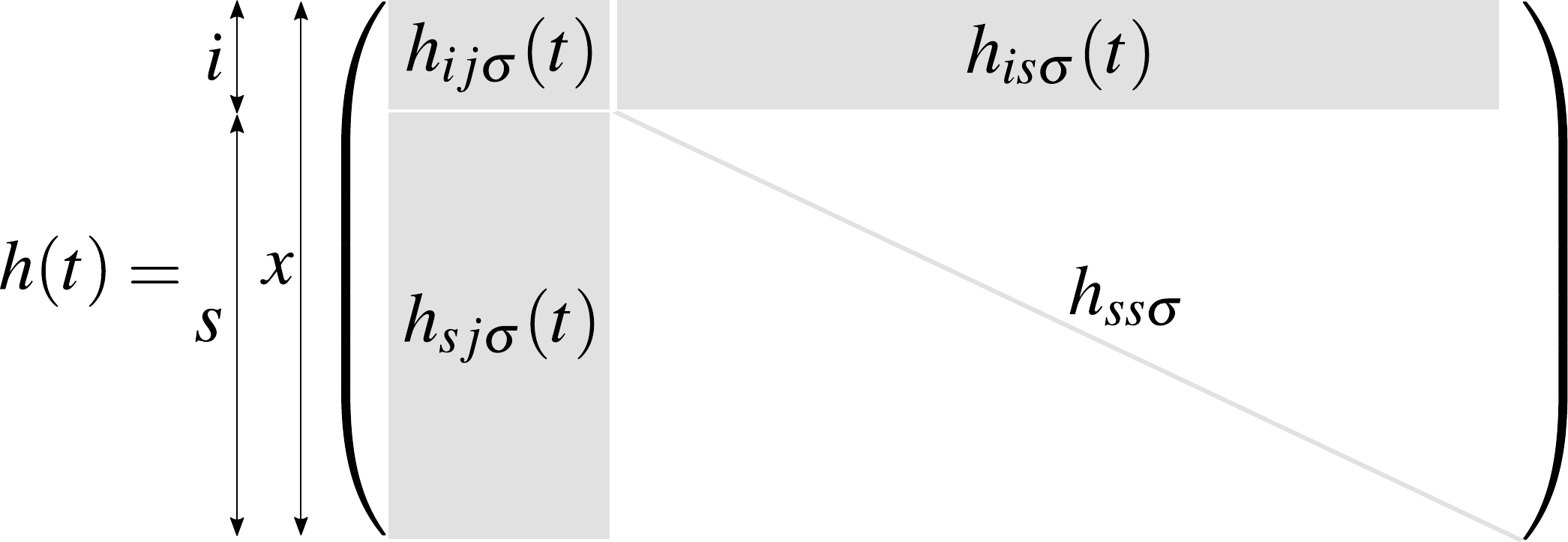}
  \caption
  {
    The effective, one-particle Hamiltonian, \eqref{eq:h_TU}, has three
    distinct kind of elements. The hybridization elements $h_{is\sigma}(t) =
    h^*_{si\sigma}(t)$, the elements $h_{ij\sigma}(t)$ of the physical sector
    and the time-independent elements $h_{ss'\sigma} \propto \delta_{ss'}$ of
    the virtual sector.
  }
  \label{fig:hcpt}
\end{figure}
The hybridization-function-like structure is the immediate and important
advantage of the Lehmann representation. It allows to write down an effective,
noninteracting model specified by the Hamiltonian
\begin{align}
  \label{eq:heff}
  H^\mathrm{eff}(t) 
  = 
  \sum_{ij\sigma} 
  &
  \left(
    T_{ij\sigma}(t) - \delta_{ij}\mu + \Sigma^\mathrm{HF}_{ij\sigma}(t)
  \right)
  \cds{i\sigma}\ccs{j\sigma}\\ 
  +
  &\sum_{is\sigma} 
  \left(
    h_{is\sigma}(t)
    \cds{i\sigma}\ccs{s\sigma} + \mathrm{h.c.}
  \right) 
  + 
  \sum_s h_{ss\sigma} \cds{s\sigma}\ccs{s\sigma}
  \nonumber , 
\end{align}
which reproduces the one-particle Green's function $G \equiv G_{T,U}$
\emph{exactly} when evaluated at the physical sites $i,j$. We emphasize that
$H^\mathrm{eff}(t)$ includes all correlation effects through the hybridization
strengths $h_{is\sigma}(t)$ and on-site energies $h_{ss\sigma}$ of the virtual
sites. Furthermore, the previously mentioned interpretation of the $s$-degrees
of freedom as additional virtual sites becomes obvious in Eq.\ \eqref{eq:heff}.

For a given arbitrary self-energy one would typically have to consider a continuum of virtual sites. 
Here, however, the situation is much simpler since $\Sigma_{ij\sigma}(t,t')$ is the CPT self-energy, i.e., the self-energy of our reference model consisting of a system of decoupled clusters. 
In this case the total number of physical and of necessary virtual sites equals the number of single-particle excitations with nonzero weight. \cite{GP15} 
As the latter grows exponentially with the size of the individual cluster, the exact mapping and the exact numerical
construction of the effective Hamiltonian is limited to clusters small enough to allow for an exact numerical diagonalization.
The computation of the parameters of the effective Hamiltonian (\ref{eq:heff}) is nontrivial but straightforward and numerically completely stable. 
Details are described in Ref.\ \onlinecite{GP15}.

For clarity, we use the following convention throughout the paper:
\begin{align} 
  \text{physical sites:}&~~i,j,k,l,\qquad\qquad\\
  \text{virtual sites:}&~~s,s',\nonumber\\ 
  \text{physical {\em or} virtual sites:}&~~x,y.\nonumber
\end{align} 
Defining $h_{ij\sigma}(t) \equiv T_{ij\sigma}(t) - \delta_{ij}\mu +
\Sigma^\mathrm{HF}_{ij\sigma}(t)$, the effective Hamiltonian can be written as
(cf.\ Fig.\ \ref{fig:hcpt})
\begin{equation}
  \label{eq:h_TU}
  H^\mathrm{eff}(t) 
  = 
  \sum_{xy\sigma} h_{xy\sigma}(t) \cds{x\sigma}\ccs{y\sigma}.
\end{equation}
The corresponding Green's function is given on the physical but also on the
virtual sites, 
\begin{equation}
  \label{eq:g_h}
  G_{xy\sigma}(t,t')
  =
  -i
  \est
  {
    \tcorder\,\cc{x\sigma}(t)\cd{y\sigma}(t')
  }_{ H^\mathrm{eff} },
\end{equation}
so that the original, physical Green's function $G_{ij\sigma}(t,t')$ is
obtained if we restrict $x,y$ to physical sites only, i.e., $(x,y) = (i,j)$.

Many-particle correlation functions, e.g., spin-spin correlations, are in
general not accessible from the effective Hamiltonian.  There is, however, one
important exception. Namely, the two-particle Green's functions $G^{(2l)}$ and
$G^{(2r)}$ can be expressed as contour convolutions of the system's self-energy
with the Green's function, cf.\ Eqs.\ \eqref{eq:g2l_def} and
\eqref{eq:g2r_def}. In the Hamiltonian-based formalism this convolution is
greatly simplified and becomes a straightforward matrix multiplication. By
comparing Dyson's equation \eqref{eq:dyson} with the equation of motion that
follows from Eq.\ \eqref{eq:g_h}, one readily finds the identity
\begin{align}
  \label{eq:tomarkovian}
  [\Sigma \circ G]_{ij\sigma}(t,t')
  =
  &\sum_{l}
  \left[
    h_{il\sigma}(t)-T_{il\sigma}(t) + \delta_{il}\mu
  \right]
  G_{lj\sigma}(t,t') \nonumber\\
  &+
  \sum_s h_{is\sigma}(t) G_{sj\sigma}(t,t').
\end{align}
An analogous relation can be derived for $G\circ \Sigma$. 

The result can be written in a more compact form by defining a new quantity
$\eta_{xy\sigma}(t)$ via
\begin{align}
  \label{eq:h=Ueta}
  h_{ij\sigma}(t)
  &=
  U(t)\eta_{ij\sigma}(t) + T_{ij\sigma}(t) - \delta_{ij}\mu,\\
  h_{is\sigma}(t)
  &=
  U(t)\eta_{is\sigma}(t),\nonumber
\end{align}
and $\eta_{ss'\sigma}(t)\equiv 0$.  This is consistent with the alternative
definition given in Appendix \ref{ap:math_eff} which also holds for $U(t) = 0$.
In the physical sector it implies
\begin{equation}
  \label{eq:eta_physical}
  \eta_{ij\sigma}(t) = \delta_{ij}\est{\hat{n}_{i\bar\sigma}(t)}_{H_{T, U}},
\end{equation}
as follows from $U(t)\eta_{ij\sigma}(t) = \Sigma^\text{HF}_{ij\sigma}(t)$ [cf.
Eqs. \eqref{eq:heff}, \eqref{eq:h_TU} and \eqref{eq:h=Ueta}].  With this
definition for $\eta$ and with the relations $\Sigma \circ G = -iU\circ
G^{(2l)}$ and $G \circ \Sigma = -iG^{(2r)}\circ U$ we get
\begin{align}
  \label{eq:g2l_hambased}
  G^{(2l)}_{ij\sigma}(t,t')
  &= 
  i\sum_{x} \eta_{ix\sigma}(t)G_{xj\sigma}(t,t'),\\
  G^{(2r)}_{ij\sigma}(t,t')
  &= 
  i\sum_{x} G_{ix\sigma}(t,t')\eta^*_{ix\sigma}(t'),
  \label{eq:g2r_hambased}
\end{align}
Recall at this point that quantities like $h(t) = h_{T,U}(t)$ or
$\eta(t) = \eta_{T,U}(t)$ (as well as $H^\mathrm{eff}(t)$, $G^{(2l)}$, etc.)
are functionals of $T$ and $U$.

\subsection{Hamiltonian-based formulation of the CPT}

Let us now discuss how the CPT Green's function can be obtained from an
effective one-particle Hamiltonian and how to set up a Markovian
time-propagation scheme. \cite{GP15} As discussed in Sec.\ \ref{sec:cpt}, we
have $T=T'+V$ where $T' - \lambda$ is the renormalized intra-cluster and $V +
\lambda$ the renormalized inter-cluster hopping.  For each set of parameters
$\lambda$, an effective one-particle CPT Hamiltonian can be defined by adding
the inter-cluster hopping to the effective Hamiltonian \eqref{eq:heff} of the
reference system:
\begin{align}
  \label{eq:hcpt}
  H^\text{CPT}[\lambda](t) 
  &=
  H^\mathrm{eff}_{T' - \lambda, U}(t)
  + 
  \sum_{ij\sigma} [V_{ij\sigma}(t) + \lambda_{ij\sigma}(t)]
  \cds{i\sigma}\ccs{j\sigma}\nonumber\\
  &\equiv 
  \sum_{xy\sigma} h^\text{CPT}_{xy\sigma}(t) \cds{x\sigma}\ccs{y\sigma}.
\end{align}
The CPT Green's function, as computed from $H^\text{CPT}[\lambda](t)$, 
\begin{equation}
  \label{eq:gcpt}
  G^\text{CPT}[\lambda]_{xy\sigma}(t,t')
  =
  -i\est{
  \tcorder
  \cc{x\sigma}(t)\cd{y\sigma}(t')
  }_{H^\text{CPT}[\lambda]}
\end{equation}
then coincides with the original definition in Eq.\ \eqref{eq:cptlambda} if
only the physical sector is considered, i.e., $(x,y)=(i,j)$. This can be
verified easily by integrating out the virtual, $s$ degrees of freedom from
$H^\mathrm{CPT}$. Eq.\ \eqref{eq:hcpt} reflects the freedom we have in the CPT
construction as the $\lambda$-terms cancel in the physical sector.  $\lambda$
only enters implicitly through the hybridization strengths $h'_{is\sigma}(t)$,
through the on-site energies $h'_{ss\sigma}$ (where $h' \equiv h_{T' - \lambda,
U}$) and through the Hartree-Fock term $\Sigma_{T'-\lambda, U}^\mathrm{HF}$ of
the reference system's Hamiltonian $H_{T' - \lambda, U}$.

For each set of parameters $\lambda$, the two-particle correlation function
$G^{(2l)}$ is approximated within the context of the CPT as 
\begin{align}
  \label{eq:g2lcpt}
  G^{(2l)}[\lambda]_{ij\sigma}(t,t') 
  &= 
  i\sum_x 
  \eta'[\lambda]_{ix\sigma}(t)
  G^\text{CPT}[\lambda]_{xj\sigma}(t,t') , 
\end{align}
where we have defined
\begin{equation}
\eta'[\lambda]\equiv \eta_{T'-\lambda, U} . 
\end{equation}
Eq.\ (\ref{eq:g2lcpt}) corresponds to the exact expression given by Eq.\
(\ref{eq:g2l_hambased}).  $G^{(2r)}[\lambda]$ is defined analogously, and thus
the symmetry relation
\begin{equation}
  \label{eq:g2symmetry}
  G^{(2r)}[\lambda]_{ji\sigma}(t,t^+) 
  = 
  \left[G^{(2l)}[\lambda]_{ij\sigma}(t,t^+)\right]^*
\end{equation}
is ensured within the CPT independently of $\lambda$. Note that this symmetry
is not sufficient to allow for an unambiguous definition of the doublon density
based on $G^{(2l)}$ and $G^{(2r)}$ [cf.\ Eq.\ \eqref{eq:d_is_independent}].
Instead, it requires both constraints to be respected as discussed in
Sec.\ \ref{sec:con_eqs}. In case of an arbitrary, non-conserving set of
parameters $\lambda$ this ambiguity needs to be circumvented by defining the
doublon density as an average 
\begin{equation}
  d_i[\lambda](t) 
  = 
  -\frac{1}{4}
  \sum_\sigma
  \left[
    G^{(2l)}[\lambda]_{ii\sigma}(t,t^+) 
    +
    G^{(2l)}[\lambda]_{ii\sigma}(t,t^+) 
  \right].
\end{equation}
For $\lambda = \lambda^\text{opt}$, however, we have
\begin{equation}
  d_i[\lambda^\text{opt}](t) 
  = 
  -G^{(2l)}[\lambda^\text{opt}]_{ii\sigma}(t,t^+) 
  = 
  -G^{(2r)}[\lambda^\text{opt}]_{ii\sigma'}(t,t^+).
\end{equation}
The final forms of the conditional equations for $\lambda^\text{opt}$ are
obtained by replacing $G^{(2r)}$ and $G^{(2l)}$ by their CPT approximations
$G^{(2l)}[\lambda]$ and $G^{(2r)}[\lambda]$ in the expressions for $F$ and $C$
given by Eqs.\ \eqref{eq:con_dens} and \eqref{eq:con_energy}:
\begin{equation}
  \label{eq:selfcon_cpt}
  F[\lambda^\text{opt}]_{i\sigma}(t)
  \stackrel{!}{=}
  0,
  \quad
  C[\lambda^\text{opt}]_{i\sigma}(t)
  \stackrel{!}{=}
  0.
\end{equation}
We note that the number of free parameters $\lambda$ must be chosen to match
the number of linear independent constraints defined by Eq.
\eqref{eq:selfcon_cpt} to ensure the existence of a unique solution
$\lambda^\text{opt}$.

\section{Solving the self-consistency equations}
\label{sec:solve_for_lambda}

Having formulated the self-consistency conditions (\ref{eq:selfcon_cpt}), it remains to explicitly solve these equations
for $\lambda^\text{opt}$. 
An important simplification arises from the fact that the CPT is by construction a fully causal theory, i.e., 
the time-local elements $G^\text{CPT} (t,t^{+})$ of the CPT Green's function at time $t$, for example, only depend on quantities at earlier times.
The same holds for $G^{(2l)}(t,t^{+})$ and for $h^\text{CPT}_{xy\sigma}(t)$.
This allows us to construct a strategy for the solution of Eq.\ (\ref{eq:selfcon_cpt}) in the form of a time-propagation algorithm. 
Let us therefore assume that $\lambda^\text{opt}$ is known for all time points on a discrete time grid and that only the parameters $\lambda^\text{opt}(t)$ at the latest point of time $t$ are unknown. 

Therewith, the actual task is to solve Eq.\ (\ref{eq:selfcon_cpt}) for $\lambda^\text{opt}(t)$ {\em only at the given latest point of time} $t$. 
To this end we have to analyze at time $t$ the $\lambda(t)$ dependence of the relevant quantities, i.e., 
of $G^{(2l)}(t,t^{+})$ and $G^{(2r)}(t,t^{+})$, see Eqs.\ (\ref{eq:con_dens}) and (\ref{eq:con_energy}).
First of all, the dependence of $G^{(2l)}(t,t^{+})$ (and $G^{(2r)}(t,t^{+})$) on $\lambda(t)$ at time $t$ is due to the CPT Hamiltonian $h^\text{CPT}_{xy\sigma}(t)$ [see Eq.\ (\ref{eq:hcpt}) and see Eqs.\ (\ref{eq:gcpt}) and (\ref{eq:g2lcpt})]. 
The $\lambda(t)$-dependence of the latter is exclusively due to the time-evolution operator $\mathcal{U}'[\lambda]\equiv \mathcal{U}_{T'-\lambda, U}$ of the reference system.
The detailed construction of $h^\text{CPT}_{xy\sigma}(t)$ is not important here, and we refer to Ref.\ \onlinecite{GP15} for a comprehensive discussion. 
Finally, the functional dependence of $U'[\lambda](t,0)$ on $\lambda$ is through an integration over all times between $0$ and $t$.
With this information at hand, we are in fact able to characterize the dependence on $\lambda(t)$ at time $t$ of the quantities $G^{(2l)}(t,t^{+})$ and $G^{(2r)}(t,t^{+})$ which enter the local constraints (\ref{eq:selfcon_cpt}) that serve to enforce the conservation laws.

The most important point for the following discussion is the fact that, in the limit of vanishing time step $\Delta t \to 0$, the parameter set $\lambda(t)$ at the latest point of time enters basically all central quantities as a null set only: 
Consider, for example, $G^{(2l)}(t,t^{+})$. 
Its first-order response due to a variation of $\lambda(t)$ at time $t$ {\em vanishes} (as shown below).
On the one hand, this missing sensitivity implies a complication of the theory since one has to account for this mathematical property explicitly when setting up a numerical implementation. 
On the other hand, once one has recognized the property, it actually helps to the solve Eqs.\ (\ref{eq:selfcon_cpt}). 
Consider a given arbitrary causal functional $M[\lambda](t)$. 
The main trick is to enhance the sensitivity of $M[\lambda](t)$ to variations of $\lambda(t)$ at time $t$ by taking its time derivative.
Typically, if the first-order response of $M[\lambda](t)$ vanishes, $\partial_{t}M[\lambda](t)$ is a {\em linear} function of $\lambda(t)$ at time $t$. 
Clearly, this is the key to solve an equation like $M[\lambda](t)=0$ for $\lambda^\text{opt}(t)$.

In the following subsections \ref{sec:a} -- \ref{sec:d} the above-sketched ideas are worked out on a more technical level. Finally, the section \ref{sec:initial_state} addresses the initial state at time $t=0$.

\subsection{Time-local variations}
\label{sec:a}

Assume that we have found the optimal renormalization $\lambda^\text{opt}(t)$
for $t \le t_n \equiv n \, \Delta t$. We introduce a variation $\deltaloc^n$
which affects the current (the $n$-th) time step only:
\begin{equation}
  \label{eq:defdeltan}
  \deltaloc^n \lambda_{ij\sigma}(t) = \delta\lambda_{ij\sigma}(t)
  \Theta^n_\mathrm{loc}(t), \quad
  \Theta^n_\mathrm{loc}(t)
  =
  \left\{
    \begin{aligned}
      1 \,\,\,\, &\text{if} \,\,\,\, {t \in \left[ t_n, t_{n+1}\right]},\\
      0 \,\,\,\, &\text{else}.
    \end{aligned}
  \right.
\end{equation}
For simplicity, we require the variations to be symmetric, i.e., $\delta
\lambda_{ij\sigma}(t) = \delta \lambda_{ji\sigma}(t)$. This implies a
restriction to symmetric solutions $\lambda^\text{opt}$. Consider now an
arbitrary, causal functional $M[\lambda](t)$, i.e., a functional that at time
$t$ only depends on $\lambda(t')$ with $t' \le t$. For such an object, the
variational operator $\deltaloc^n$ is related to the conventional functional
derivative through
\begin{align}
  \deltaloc^n M[\lambda](t) 
  &=
  \sum_{\sigma}\sum_{i \ge j}
  \int_{t_n}^t \dt'
  \frac{\delta M[\lambda](t)}{\delta \lambda_{ij\sigma}(t')}
  \delta\lambda_{ij\sigma}(t'),
\end{align}
where the restriction $i\ge j$ is necessary because of the symmetry requirement
$\lambda_{ij\sigma}=\lambda_{ji\sigma}$. 

We now take the combined limit $n\rightarrow\infty,\, \Delta t \rightarrow 0$ 
such that we always have $t \in [t_n, t_{n+1}]$ to define the time-local
variation $\deltaloc$ in the continuum limit 
\begin{align}
  \deltaloc M[\lambda](t)
  &= 
  \lim_{\substack{\Delta t \rightarrow 0 \\ n\rightarrow \infty}}
  \deltaloc^n M[\lambda](t),
\end{align}
with the corresponding variational quotient
\begin{align}
  \frac{\deltaloc M[\lambda](t)}{\deltaloc \lambda_{ij\sigma}(t)}
  \equiv
  \lim_{\substack{\Delta t \rightarrow 0 \\ n\rightarrow \infty}}
  \int_{t_n}^t \dt'
  \frac{\delta M[\lambda](t)}{\delta \lambda_{ij\sigma}(t')}.
\end{align}
This variational quotient describes the linear response of $M[\lambda](t)$ when
varying the parameters at the latest time step:
\begin{align}
  \deltaloc M[\lambda](t)
  =
  \sum_{\sigma}\sum_{i \ge j}
  \frac{\deltaloc M[\lambda](t)}{\deltaloc \lambda_{ij\sigma}(t)}
  \delta\lambda_{ij\sigma}(t).
\end{align}

\subsection{Integrated quantities in $\lambda$}
\label{sec:b}

With the appropriate variation for our purposes at hand, we can study the
effect of the variation on the main quantities within the CPT framework.  We
first consider the time-evolution operator (``propagator'') of the reference
system $\mathcal{U}'[\lambda]\equiv \mathcal{U}_{T'-\lambda, U}$.  It is
instructive to study the effect of the operator $\deltaloc^n$ first, i.e., the
effect of a time-local variation with finite time step $\Delta t$. Keeping only
terms of the order $O(\Delta t)$ one finds
\begin{align}
  \deltaloc^n \mathcal{U}'[\lambda](t,0)
  &= 
  -i 
  \left[
    \sum_{ij\sigma} 
    \int_{t_n}^{t}
    \delta\lambda_{ij\sigma}(t') \cd{i\sigma}\cc{j\sigma} \dt'
  \right]
  \mathcal{U}'[\lambda](t_n,0)\nonumber\\
  &\phantom{=}\,\,+ 
  O(\Delta t^2).
  \label{eq:dU_vanishes}
\end{align}
In lowest order we thus have $\deltaloc^n \mathcal{U}_{T' - \lambda,U}(t,0)
\propto \Delta t \, \delta \lambda(t)$.  This means that the linear response
vanishes identically in the limit $\Delta t \rightarrow 0$. This property
originates from the fact that $\lambda(t)$ is integrated over time within the
propagator $\mathcal{U}_{T' - \lambda, U}(t,0)$, and that the contribution of a
single time step, $t \in [t_n, t_{n+1}]$, to this integral is of zero measure
in the limit $\Delta t \rightarrow 0$. 

A finite time-local variation is obtained for the first time derivative
of the propagator rather than for the propagator itself. 
Namely, the corresponding time-local variational quotient remains
non-zero in the continuum limit:
\begin{equation}
  \label{eq:dU_explicit}
  \frac{\deltaloc [i\partial_t \mathcal{U}'[\lambda](t,0)]}
  {\deltaloc\lambda_{ij\sigma}(t)}
  = 
  -
  \bigl[
    \cds{j\sigma}\ccs{i\sigma } 
    + 
    \cds{i\sigma}\ccs{j\sigma} 
    -
    \delta_{ij}\cds{i\sigma}\ccs{i\sigma}
  \bigr]
  \mathcal{U}'[\lambda](t,0).
\end{equation}
Multiplying this equation with $\lambda_{ij\sigma}(t)$, summing over
$i,j,\sigma$ and comparing with the standard equation of motion $i\partial_t
\mathcal{U}'[\lambda](t,0) = H_{T' - \lambda, U}(t)
\mathcal{U}'[\lambda](t,0)$, shows that the time derivative of the propagator
is of the general form
\begin{equation}
  \label{eq:dU_form}
  i\partial_t \mathcal{U}'[\lambda](t,0)
  =
  \sum_\sigma\sum_{i\ge j}
  \frac{\deltaloc [i\partial_t \mathcal{U}'[\lambda](t,0)]}
  {\delta_\mathrm{loc}\lambda_{ij\sigma}(t)}
  \lambda_{ij\sigma}(t)
  +
  \xi_{\mathcal{U}'}[\lambda](t),
\end{equation}
where $\xi_{\mathcal{U}'}[\lambda](t) = H_{T', U}(t)
\mathcal{U}'[\lambda](t,0)$.
Note that the dependence on $\lambda_{ij\sigma}(t)$ at time $t$ is strictly {\em linear}
in the limit $\Delta t\to 0$.

With this definition and with Eq.\ \eqref{eq:dU_explicit}, it is obvious that 
the variational derivative  and $\xi_{\mathcal{U}'}[\lambda](t)$ on the
right-hand side of Eq.\ (\ref{eq:dU_form}) depend on $\lambda(t)$ only through
an integration over time within the propagator $\mathcal{U}'[\lambda](t,0)$.
We will call such quantities \emph{integrated} quantities in $\lambda$.
Integrated quantities in $\lambda$ inherit an important property from the
cluster propagator $\mathcal{U}'[\lambda]$, see Eq.\ \eqref{eq:dU_vanishes}:
Their time-local variation vanishes in the limit $\Delta t \rightarrow 0$. 

Furthermore, the time derivative of any quantity $M[\lambda]$ that is
integrated in $\lambda$, i.e., the time derivative of a functional of the form
$M[\lambda](t) = M(\mathcal{U}'[\lambda](t,0))$,
can be brought into a form analogous to Eq.\ \eqref{eq:dU_form}. This follows
immediately from the chain rule in calculus as $i\partial_t M[\lambda](t) =
\frac{i\partial M(\mathcal{U}')}{\partial \mathcal{U}'}\frac{\partial
\mathcal{U}'[\lambda](t,0)}{\partial t}$. Explicitly this result reads
\begin{equation}
  \label{eq:dM_form}
  i\partial_t M[\lambda](t)
  =
  \sum_\sigma\sum_{i\ge j}
  \frac{\deltaloc [i\partial_t M[\lambda](t)]}
  {\delta_\mathrm{loc}\lambda_{ij\sigma}(t)}
  \lambda_{ij\sigma}(t)
  +
  \xi_{M}[\lambda](t),
\end{equation}
where $\frac{\deltaloc [i\partial_t M[\lambda](t)]}
{\delta_\mathrm{loc}\lambda_{ij\sigma}(t)}$ and $\xi_M[\lambda]$ are again
integrated quantities in $\lambda$. We furthermore conclude that a time-local
variation of the time derivative of an integrated quantity in $\lambda$ is
non-zero in general.

The main idea in the following is to combine the conditional equations
\eqref{eq:selfcon_cpt} into a single equation $\Gamma[\lambda^\text{opt}](t)
\stackrel{!}{=} 0$ such that $\Gamma[\lambda]$ is of the form
$\Gamma[\lambda](t) = J[\lambda](t) \lambda(t) + \xi_\Gamma[\lambda](t)$ where 
$J[\lambda]$ and $\xi_\Gamma[\lambda]$ are integrated quantities in $\lambda$.
This is formally easily solved for $\lambda^\text{opt}(t)$ by matrix inversion
and allows to derive an efficient propagation scheme for numerical purposes.

\subsection{$\lambda$-dependence of $G^{(2l)}$ and $G^{(2r)}$}
\label{sec:c}

The main building blocks of the local constraints on the spin-dependent
density, Eq.\  \eqref{eq:con_dens}, and the doublon density, Eq.
\eqref{eq:con_energy}, are given by the two-particle correlation functions
$G^{(2l)}$ and $G^{(2r)}$. Within the CPT approximation they are defined
through Eq.\  \eqref{eq:g2lcpt}. We therefore have to understand the $\lambda$
dependence of $\eta'[\lambda]\equiv \eta_{T' - \lambda, U}$ and
$G^\text{CPT}[\lambda]$.

One can easily see that $\eta'[\lambda]$ is an integrated quantity in
$\lambda$.  Consider, for example, the physical sector.  From Eq.\
\eqref{eq:eta_physical} we have $\eta'_{ij\sigma}[\lambda](t) =
\delta_{ij}\est{\hat{n}_{i\bar\sigma}(t)}_{H_{T' - \lambda, U}}$. The only
$\lambda$-dependence of this expression indeed stems from the propagator
$\mathcal{U}'[\lambda]$. To obtain a non-vanishing time-local variation we thus
have to consider the first derivative with respect to time.  This is worked out
in Appendix \ref{ap:d_eta}:
\begin{align}
  \deltaloc[i\partial_t \eta'_{ij\sigma}(t)]
  &
  =
  \eta'_{ii\sigma}(t)\delta\lambda_{ij\sigma}(t)
  -
  \sum_{l\sigma'}
  [\delta\lambda_{il\sigma'}(t)]\gamma^{l\sigma'}_{ij\sigma}(t),
  \nonumber \\
  \deltaloc[i\partial_t \eta'_{is\sigma}(t)]
  &
  =
  -\sum_{l\sigma'}
  [\delta \lambda_{il\sigma'}(t)]\gamma^{l\sigma'}_{is\sigma}(t),
  \label{eq:eta_variation}
\end{align}
where the newly introduced tensor $\gamma[\lambda]^{l\sigma'}_{is\sigma}(t)$ is
cluster-diagonal, i.e., $\gamma[\lambda]^{l\sigma'}_{is\sigma}(t) \neq 0$ if
and only if $i$ and $l$ refer to lattice sites within the same cluster.  It
furthermore follows that $i\partial_t \eta'[\lambda](t)$ can be brought into
the form specified by Eq.\ \eqref{eq:dM_form}, where the variational derivative
$\frac{\deltaloc [i\partial_t \eta'[\lambda]_{ix\sigma}(t)]}
{\delta\lambda_{jl\sigma}(t)}$, as given by Eq.\ \eqref{eq:eta_variation}, and
$\xi_{\eta'}[\lambda]_{ix\sigma}(t)$ are integrated quantities in $\lambda$.
An explicit expression for the latter is not needed for our purposes.

Let us now take a look at the CPT Green's function. It depends on $\lambda$
through the Hamiltonian $H^\mathrm{CPT}[\lambda]$, which in turn depends on
$\lambda$ through the hybridization strengths $h'[\lambda]_{is\sigma}(t) = U(t)
\eta'[\lambda]_{is\sigma}(t)$ and the Hartree-Fock term
$\Sigma'[\lambda]^\text{HF}_{ij\sigma}(t) = U(t)\eta'[\lambda]_{ij\sigma}(t)$.
The Hamiltonian $H^\text{CPT}(t)$ is therefore an integrated quantity in
$\lambda$. As the propagator $\mathcal{U}^\text{CPT}[\lambda](t,0) = T
\exp{-i\int_0^t \dt' H^\text{CPT}[\lambda](t')}$ involves a second integral
over time, we conclude that $\deltaloc^n G^\text{CPT}(t,t^+) \propto \Delta
t^2 \, \delta \lambda(t)$. In this sense, $G^\text{CPT}$ must be seen as an
integrated quantity in $\lambda$ of second order. Consequently, the time-local
variation of its first derivative with respect to time vanishes: 
\begin{equation}
  \label{eq:gcpt_vanishes}
  \deltaloc [i\partial_t G^\text{CPT}[\lambda](t,t^+)]
  = 0.
\end{equation}
We note that the time derivative involves the product of the matrix elements of
the CPT Hamiltonian, Eq.\ (\ref{eq:hcpt}), with $G^\text{CPT}[\lambda](t,t^+)$,
i.e., the product of an integrated quantity in $\lambda$ with an integrated
quantity in $\lambda$ of second order, respectively.  Obviously, the product
scales like an ordinary integrated quantity in $\lambda$ when a time-local
variation is applied, i.e., $\deltaloc^n h^\text{CPT}(t) G^\text{CPT}(t,t^+)
\propto \Delta t \delta\lambda(t)$ in lowest order in $\Delta t$.

Concluding, to get a non-vanishing time-local variation, one must consider the
first time derivative of the two-particle Green's functions $G^{(2l)}$ and
$G^{(2r)}$.  We find
\begin{align}
  \label{eq:deltaG}
  \deltaloc 
  \left[
    i \partial_t G^{(2l)}_{ij\sigma}(t,t^+) 
  \right]
  =
  \sum_{x}
  \left(
    \deltaloc [i\partial_t \eta'_{ix\sigma}(t)]
  \right)
  G^\text{CPT}_{xj\sigma}(t,t^+)
\end{align}
and an analogous expression for $G^{(2r)}$.  Only the $\eta'$-term contributes
to the variation, cf.\ Eq.\ \eqref{eq:eta_variation}, while the variation of
the CPT Green's function vanishes, cf.\ Eq.\ \eqref{eq:gcpt_vanishes}.  We also
note that Eq.\ \eqref{eq:eta_variation} may be used at this point and that
$i\partial_t G^{(2l)}(t,t^+)$ [and analogously $i\partial_t G^{(2r)}(t,t^+)$]
is of the form
\begin{equation}
  \label{eq:idtG2}
  i\partial_t G^{(2l)}_{ij\sigma}(t,t^+)
  =
  \sum_{\sigma'}\sum_{k\ge l}
  \frac{\deltaloc G^{(2l)}_{ij\sigma}(t,t^+)}{\deltaloc \lambda_{kl\sigma'}(t)}
  \lambda_{kl\sigma'}(t)
  +
  [\xi_{G^{(2l)}}]_{ij\sigma}(t),
\end{equation}
where both, the variational derivative $\frac{\deltaloc G^{(2l)}(t,t^+)}
{\deltaloc \lambda(t)}$, as given by Eq.\ \eqref{eq:deltaG}, and the quantity
$\xi_{G^{(2l)}}(t)$, which is not needed in explicit form for our purposes,
scale like integrated quantities in $\lambda$ under time-local variations. This
follows from the fact that $i\partial_t G^\text{CPT}(t,t^+)$ scales like an
integrated quantity in $\lambda$ under time-local variations and the related
discussion above.

\subsection{$\lambda$-dependence of the local constraints} 
\label{sec:d}

The local constraint on the spin-dependent density, Eq.\
\eqref{eq:con_dens}, is formulated in terms of the difference between
$G^{(2l)}$ and $G^{(2r)}$. Therefore its first derivative with respect to time
must be considered to obtain a non-vanishing time-local variation:
\begin{align}
  \label{eq:var_F}
  \deltaloc [i\partial_t F[\lambda]_{i\sigma}(t)]
  =
  \deltaloc 
  &\left[
  i\partial_t G^{(2l)}[\lambda]_{ii\sigma}(t,t^+) 
  \right.\\
  &\quad\quad\quad\quad \left.
    -
    i\partial_t G^{(2r)}[\lambda]_{ii\sigma}(t,t^+) 
  \right],\nonumber
\end{align}
For the time-local variation of the local constraint on the doublon density,
Eq.\ \eqref{eq:con_energy}, on the other hand, one finds
\begin{align}
  \label{eq:var_C}
  \deltaloc C[\lambda]_{i\sigma}(t)
  =
  \deltaloc 
  &\left[
  i\partial_t G^{(2l)}[\lambda]_{ii\sigma}(t,t^+) 
  \right.\\
  &\quad\quad\quad\quad \left.
    +
    i\partial_t G^{(2r)}[\lambda]_{ii\sigma}(t,t^+) 
  \right],\nonumber
\end{align}
since $\deltaloc [T \circ G^{(2r)} - G^{(2l)} \circ T](t,t^+) = 0$, where we
made use of the fact that $T = T' + V$ is the hopping of the original system
and thus independent of $\lambda$.

To treat both constraints in a combined formal frame, we define 
the functional $\Gamma[\lambda]$:
\begin{equation}
  \label{eq:DefG}
  \Gamma[\lambda]_a(t) = 
  \left\{ 
    \begin{aligned}
      &i \partial_t
       F[\lambda]_{i\sigma}(t) \quad \text{if} \quad 0 \le a < 2L\: , \\
      &C[\lambda]_{i\sigma}(t) \quad \text{if} \quad 2L \le a < 4L \: , 
    \end{aligned}
  \right.
\end{equation}
where $L$ is the number of lattice sites. With this, the conditional
equation for the optimal renormalization reads $\Gamma[\lambda^\text{opt}]
\stackrel{!}{=} 0$. From the previous discussion and Eq.\ \eqref{eq:idtG2} it
follows that $\Gamma[\lambda]_a(t)$ is of the form
\begin{equation}
  \label{eq:Gform}
  \Gamma[\lambda]_a(t) 
  = 
  \sum_{b}  J[\lambda]_{ab}(t) \lambda_b(t)
  +
  \xi_\Gamma[\lambda]_a(t),
\end{equation}
where we introduced the super-index $b$ which labels the set of free
parameters: $\lambda_b = \lambda_{ij\sigma}$, $i \ge j$. Both $J[\lambda]$ and
$\xi_\Gamma[\lambda]$ scale like integrated quantities in $\lambda$ under
time-local variations. The Jacobian matrix $J$ is defined as 
\begin{equation}
  J[\lambda]_{ab}(t) 
  \equiv 
  \frac{\deltaloc \Gamma_a[\lambda](t)}{\deltaloc \lambda_b(t)}.
\end{equation}
The matrix $J[\lambda](t)$ is quadratic if the number of free parameters
$\lambda_b$ is chosen such that it equals the number of conditional equations
[see Eq.\ \eqref{eq:DefG}].  Assuming that $J[\lambda](t)$ is regular, one can
formally solve the conditional equation for the optimal renormalization:
\begin{equation} 
  \label{eq:lopt}
  \Gamma[\lambda^\text{opt}](t) \stackrel{!}{=} 0
  \quad
  \Leftrightarrow
  \quad
  \lambda^\mathrm{opt}(t) 
  = 
  -
  \left[
    J[\lambda^\text{opt}](t)
  \right]^{-1}
  \xi_\Gamma[\lambda^\text{opt}](t) \: , 
\end{equation}
see Eq.\ (\ref{eq:Gform}). This completes our derivation. 

Let us emphasize that the single point $\lambda^\mathrm{opt}(t)$ represents a
null set with respect to the time-integrations in $J[\lambda^\text{opt}](t)$
and $\xi_\Gamma[\lambda^\text{opt}](t)$. This can be exploited to derive an
efficient numerical scheme to obtain $\lambda^\text{opt}(t)$ step by step on
the time axis as detailed in Appendix \ref{sec:high-order}. There we also
argue why finding an explicit expression for $\xi_\Gamma[\lambda](t)$ can in
fact be circumvented.  An explicit expression for $J[\lambda](t)$ in terms of
known quantities, on the other hand, is available via Eqs.
\eqref{eq:eta_variation}, \eqref{eq:deltaG}, \eqref{eq:var_F} and
\eqref{eq:var_C}. 

\subsection{The equilibrium initial state}
\label{sec:initial_state}

Initially, at time $t=0$ the system is assumed to be in a thermal state.  The
CPT approximation for the initial thermal state suffers from the fact that the
starting point of the all-order perturbation theory in the inter-cluster
hopping is not unique.  This is completely analogous to the CPT description of
the real-time dynamics.  Unlike the real-time dynamics, however, the local 
constraints cannot be used to fix the renormalization parameters
$\lambda_\mathrm{eq}\equiv \lambda(0)$ for the initial state, and thus a
nontrivial self-consistency condition is not available, unfortunately.

This can be seen as follows: Let us assume that the hopping matrix $T$, and
consequently $T'$ and $V$, are real and symmetric.  Consider $G^{(2l)}$ at
times $t=t'=0$.  Via Eq.\ \eqref{eq:g2lcpt} this is given as
$G^{(2l)}[\lambda]_{ij\sigma}(0,0^+) = i\sum_{x} \eta'[\lambda]_{ix\sigma}(0)
G^\text{CPT}[\lambda]_{xj\sigma}(0,0^+)$.  In Appendix \ref{ap:math_eff} it is
shown that at time $t=0$ the imaginary part $\mathrm{Im}\,\{
\eta'[\lambda]_{ix\sigma}(0)\}$ vanishes, independently of
$\lambda_\mathrm{eq}$.  Hence, Eq.\  \eqref{eq:h=Ueta} implies that
$H^\text{CPT}[\lambda](0)$ is real and symmetric, and therefore
$G^\text{CPT}[\lambda]_{xy\sigma}(0,0^+) = i\est{\cd{y\sigma}(0)
\cc{x\sigma}(0)}_{H^\text{CPT}[\lambda]}$ is purely imaginary.  Consequently,
$G^{(2l)}[\lambda](0,0^+)$ is real.  Finally, we conclude with Eq.\
\eqref{eq:g2symmetry} that
\begin{equation}
  G^{(2l)}[\lambda]_{ij\sigma}(0,0^+) = G^{(2r)}[\lambda]_{ji\sigma}(0,0^+).
\end{equation}
This directly proves that $F[\lambda]_{i\sigma}(0)=0$, and furthermore
\begin{align}
  C[\lambda]_{i\sigma}(0)
  =
  2\sum_{j\sigma}
  T_{ij\sigma}(0)
  &\left[
    G^{(2r)}[\lambda]_{ji\sigma}(0,0^+)
  \right.\\
  &\quad\quad
  \left.
    -
    G^{(2l)}[\lambda]_{ij\sigma}(0,0^+)
  \right]
  =
  0
  \nonumber 
\end{align}
irrespective of $\lambda_\mathrm{eq}$, i.e., both constraints hold
trivially. 

For the concrete numerical
calculations we therefore circumvent this issue and consider a noninteracting initial
state. In this case the CPT is exact, independent of the choice of $\lambda$.
The initial value $\lambda_\mathrm{eq}$ is then fixed by requiring $\lambda$ to
be continuous so that $\lambda_\mathrm{eq} = \lambda(0^+)$. 

\section{Numerical results}
\label{sec:num}

\begin{figure}
\includegraphics[width=0.16\textwidth]{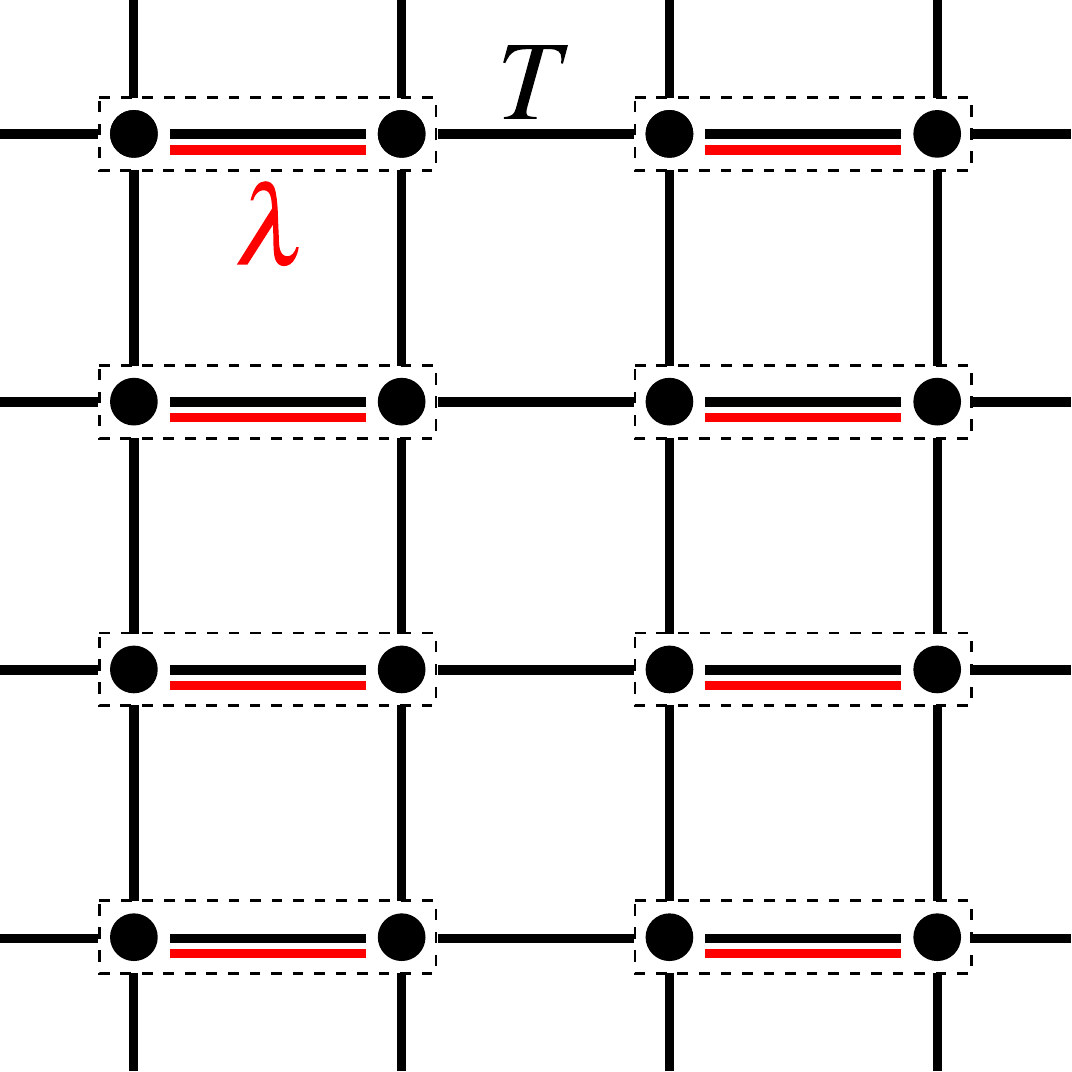}
\caption{
  Partitioning of the Hubbard model with nearest-neighbor hopping on the
  two-dimensional square lattice used for the numerical calculations. Original
  system: $L=10 \times 10$ lattice with periodic boundary conditions.
  Reference system: $50$ clusters of size $2 \times 1$.  The figure shows a $4
  \times 4$ excerpt.  Clusters are indicated by dashed rectangles.
  Nearest-neighbor hopping $T$ and optimization parameter $\lambda(t)$ are
  indicated by black and red lines.  The time-dependent renormalization
  $\lambda(t)$ is employed to enforce the conservation of energy in the
  real-time dynamics following an interaction quench.
}
\label{fig:setup}
\end{figure}

The conserving CPT has been implemented numerically.  First results are
discussed for the two-dimensional Hubbard model on an $L=10 \times 10$ square
lattice with periodic boundary conditions.  As these results shall serve as a
proof of concept only, we restrict ourselves to the most simple approximation,
i.e., to the smallest meaningful cluster as the building block of the reference
system, namely a cluster consisting of $2 \times 1$ sites.  Hence, the entire
system is partitioned into $50$ clusters in total, see Fig.\ \ref{fig:setup}.

Initially, the system is prepared in its noninteracting ground state at
half-filling by choosing $\mu = 0$.  Note that the CPT description of this
initial state is exact (and independent of the renormalization).  The hopping
of the original system is restricted to nearest neighbors, and we set the
nearest-neighbor hopping $T=1$ to fix energy and time units.  To drive the
system out of equilibrium we consider an interaction quench where the
Hubbard-$U$ is suddenly, at time $t=0$, switched on to a finite value
$U_{\text{fin}}$: 
\begin{equation} 
  U(t)=\Theta(t) U_\text{fin}.  
\end{equation}
Here, $\Theta(t)$ denotes the Heaviside step function.  For times $t > 0$ the
interaction strength is constant.  To maintain particle-hole symmetry and
half-filling, the chemical potential is quenched as well, from $\mu=0$ to
$\mu= U/2$ in the final state.

Studying the model at the particle-hole symmetric point is convenient since the
conservation of the total particle number is trivially respected in this case.
\cite{JP13} For a spin-independent parameter quench, as considered here, the
CPT also trivially respects the conservation of the total spin.  Total-energy
conservation, on the other hand, is violated in a conventional CPT approach as
has been explicitly demonstrated recently. \cite{GP15} For the present setup we
will therefore employ the nearest-neighbor hopping within the $2 \times 1$
reference system to enforce the energy-conservation law.  This specifies the
time-dependent renormalization parameter $\lambda(t)$ (see Fig.\
\ref{fig:setup}).

We note that the computational effort to self-consistently evaluate the presented theory numerically is essentially determined by the underlying solver for the conventional nonequilibrium CPT with little overhead. 
Here, we use the time-local, Hamiltonian-based solver developed in Ref.\ \onlinecite{GP15} which constructs
the effective Hamiltonian of each cluster by exact diagonalization. 
For the $2 \times 1$ reference system under consideration, only two virtual sites are needed for an exact mapping. This gives us four sites per cluster so that the CPT-Hamiltonian comprises $200$ sites in total. 
Furthermore, regarding computational demands, our approach inherits a constant memory consumption from the CPT solver as well as the linear scaling in the maximum propagation time.
In particular, we have used a time step of $\Delta t = 0.001$ to propagate the system up to $26,500$ steps up to a maximum propagation time of $t_\text{max}=26.5$. For each such time step $t_n$, the scheme developed in Appendix \ref{sec:high-order} has been employed with $n_p=1$, i.e., we have calculated the Taylor coefficients $\lambda_{n,0}$ and $\lambda_{n,1}$.

While the required computational resources are very moderate, accessing longer
time scales has turned out to be hindered by mathematical complications.  As is
obvious from Eq.\ \eqref{eq:lopt}, an inversion of the Jacobian matrix $J(t)$
is necessary to obtain $\lambda^\text{opt}(t)$ at each time step.  However,
with increasing $U_\text{fin}$ this matrix exhibits singular points of
non-invertibility at earlier and earlier times.  In fact, one finds numerically
that also the starting point $t=0^+$ is singular, namely the Jacobian matrix
vanishes: $J(0^+)=0$.  Fortunately, one also has $\xi_\Gamma(0^+)=0$, such that
this problem is fixed by applying L'H\^{o}pital's rule.  At time $t=0^+$, the
defining equation for $\lambda^\text{opt}(0^+)$ becomes
\begin{equation}
  \lambda^\mathrm{opt}(0^+) 
  = 
  -
  \left[
    \partial_t J[\lambda^\text{opt}](0^+)
  \right]^{-1}
  [\partial_t \xi_\Gamma[\lambda^\text{opt}](0^+)].
\end{equation}
While this solves the problem at time $t=0^{+}$, finding a systematic and
convenient way to propagate beyond the singular points of the Jacobian matrix
at {\em finite} times remains topic for future investigations.

\begin{figure}
\includegraphics[width=0.45\textwidth]{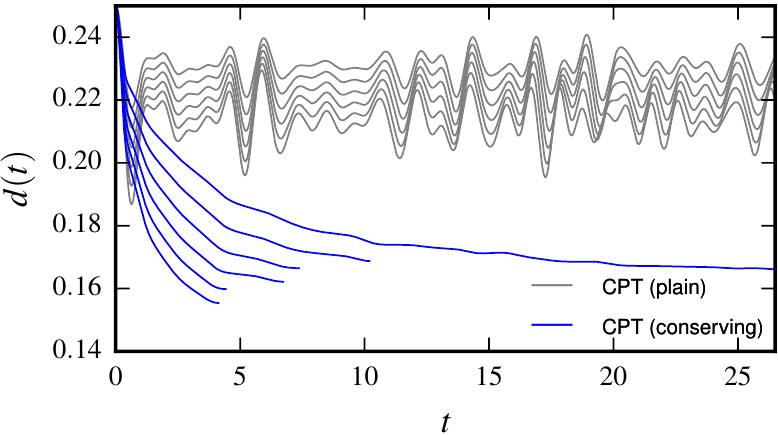}
\caption{
  Time evolution of the local doublon density after an interaction quench
  $U=0\rightarrow U_\text{fin}$.  Grey lines: plain CPT.  Blue lines:
  conserving CPT.  Results for different $U_\text{fin}$ ranging from
  $U_\text{fin} = 0.5$ (top curve) to $U_\text{fin} = 1.0$ (bottom) with
  equidistant steps $\Delta U_\text{fin} = 0.1$.  For the conserving CPT
  calculations, propagation times are limited by singular Jacobians.
}
\label{fig:docc}
\end{figure}

Apart from this technical problem, the suggested scheme works as expected.
Results for the time evolution of the doublon density are shown in Fig.\
\ref{fig:docc}.  It is evident that the renormalization $\lambda$ has a strong
influence on the dynamics and leads to qualitatively different results when
comparing the plain unoptimized CPT calculation with the novel conserving CPT.
While the dynamics is characterized by ongoing oscillations when using plain
CPT, there is a monotonic decay of the doublon density in case of the
conserving CPT.  The longest maximum propagation time is achieved for the
quench $U = 0\rightarrow 0.5$.  Here, the first singular point of the Jacobian
shows up at $t_\text{max} \approx 26.5$.  On this time scale, the doublon
density seems to relax to a stationary state with little to no oscillations. 

\begin{figure}
\includegraphics[width=0.45\textwidth]{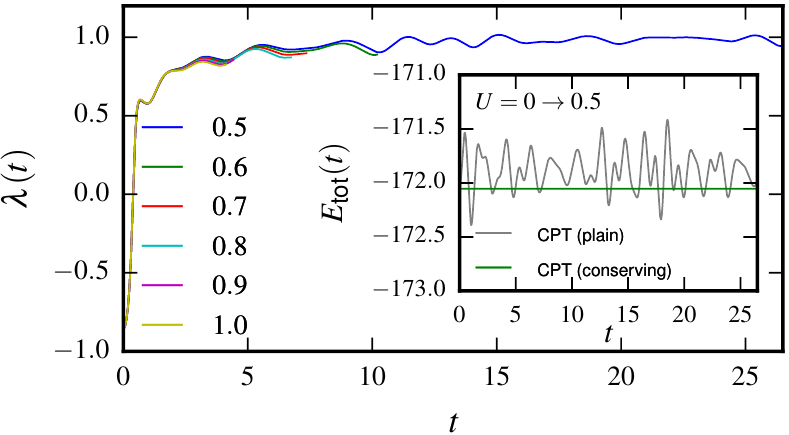}
\caption{
  Time evolution of the optimal renormalization parameter
  $\lambda^\text{opt}(t)$ for different $U_\text{fin}$ as indicated and
  corresponding to Fig.\ \ref{fig:docc}.  Inset: time dependence of the total
  energy (plain and conserving CPT).
}
\label{fig:energy}
\end{figure}

The qualitatively different time dependence of the doublon density reflects the
qualitatively different behavior found for the total energy in the plain and
the conserving CPT: This is shown in the inset of Fig.\ \ref{fig:energy}.  For
the conserving CPT, the total energy is perfectly conserved within numerical
accuracy -- by construction of the approach.  In the plain CPT calculation,
however, the total energy shows unphysical oscillations.  Here, maxima and
minima of $E_{\rm tot}(t)$ nicely correspond to maxima and minima in the
plain-CPT doublon density seen in Fig.\ \ref{fig:docc}.  It must be concluded
that those are artifacts of the plain CPT approach.  We also note that small
unphysical oscillations of the total energy {\em density} $E_{\rm tot}(t) / L$
(with $L=100$) with amplitudes less than $0.01$ lead to much stronger
oscillations in the doublon density with amplitudes of about 0.04.

The main part of Fig.\ \ref{fig:energy} displays the results for the time
evolution of the renormalization parameter $\lambda(t)$.  Its dependence on
$U_\text{fin}$ turns out to be rather weak on a time scale of a few inverse
hoppings.  Irrespective to the final interaction strength $U_{\rm fin}$, the
initial equilibrium value is found as $\lambda_\mathrm{eq} \approx -0.86$.  For
$t>0$ and for all $U_{\rm fin}$, the renormalization parameter rapidly
increases to $\lambda \approx 0.6$ within a very short time $t \approx 0.7$.
This corresponds to the rapid initial drop of the doublon density (cf.\ Fig.\
\ref{fig:docc}).  Results for longer times are again only available for the
quench $U = 0 \rightarrow 0.5$.  On the time scale up to $t_{\rm max} \approx
26.5$, we observe a subsequent slow relaxation of $\lambda(t)$ toward an
average final value $\lambda_\infty \approx 1.0$ with small superimposed
oscillations.  It seems reasonable to assume that a similar behavior would also
be found for the other quenches, given that the short-time dynamics is very
similar for the different $U_{\rm fin}$.

One should note that $\lambda_\infty = 1$ amounts $T' - \lambda_{\infty} = 0$,
i.e., a vanishing renormalized intra-cluster hopping.  Apart from the remaining
oscillations of the renormalization parameter around $\lambda_\mathrm{\infty} =
1$, this means that the system ``chooses'' the atomic limit of the Hubbard
model as the optimal starting point for the all-order perturbation theory in
the inter-cluster hopping for long times.  This may be interpreted as follows:
First of all, the remaining oscillations are understood as being necessary to
keep the total energy constant within the conserving CPT.  Disregarding the
oscillations, the value $\lambda_\mathrm{\infty} = 1$ means that, on the level
of the reference system, the doublon density becomes a conserved quantity for
long times.  This, however, is in fact a plausible starting point if the
doublon density of the full lattice model approaches a constant in the course
of time.  As is seen in Fig.\ \ref{fig:docc}, this is almost the case.  The
remaining time dependence of the doublon density of the lattice model is weak
and would be exclusively due to the inter-cluster hopping (if $\lambda(t) =
\lambda_\mathrm{\infty} = 1$ exactly).  

Let us compare the CPT result for the doublon density with the results of
previous calculations for the one-dimensional Hubbard model \cite{RP17} using
the density-matrix renormalization group (DMRG) and for the model in infinite
dimensions using the dynamical mean-field theory (DMFT). \cite{EKW09} In both
cases, a very fast relaxation of the doublon density on a time scale of one
inverse hopping has been found in fact.  Typically, however, the doublon
density first develops a minimum before it saturates to an almost constant
value.  This minimum is absent in the conserving CPT calculations (see Fig.\
\ref{fig:docc}).  Note, that for weak quenches and on the intermediate time scale discussed here and in the DMRG and DMFT studies, the doublon density does not relax to its thermal value due to kinematic constraints becoming active after the ultrashort initial relaxation step. \cite{MK08,KPD+17} 
Indeed, one expects a subsequent relaxation on a much longer time scale.  
Let us emphasize that while our data in Fig.\ \ref{fig:docc} are compatible with these expectations, serious predictions
using the conserving CPT are not yet possible.  
This would require a much more systematic study involving different and in particular larger clusters, an
analysis of the dependence on the cluster shape and also a systematic
discussion of the different possibilities to choose renormalization parameters
for the self-consistent procedure.

\section{Conclusions}
\label{sec:con}

Cluster-perturbation theory, as proposed originally, represents the most simple
way to construct a mean-field theory which incorporates to some extent the
effects of short-range correlations.  We have emphasized that the starting
point of the perturbational expansion in the inter-cluster hopping is by no
means predetermined and that the according freedom in the choice of the
intra-cluster hopping parameters can be exploited to ``optimize'' the
mean-field theory.  There are different conceivable optimization schemes.  One
way is to add an additional self-consistency condition as, for example, a
self-consistent renormalization of the on-site energies which would be very
much in the spirit of the Hubbard-I approximation.  The disadvantage of such
ideas is their {\em ad hoc} character.  An optimization following a general
variational principle is much more satisfying and physically appealing.  This
is the route that is followed up by self-energy-functional theory.
Unfortunately, total-energy conservation is not straightforwardly implemented
within the SFT context.  An appealing idea is thus to use the above-mentioned
freedom to {\em enforce} energy conservation, and actually any conservation law
dictated by the symmetries of the problem at hand.  This leads to the
conserving CPT proposed in the present paper. 

As we have argued (see Sec.\ \ref{sec:initial_state}), this idea can
exclusively be used to constrain the CPT real-time dynamics while other
concepts must be invoked for the initial thermal state.  On the other hand,
there is an urgent need for numerical approaches, even for comparatively simple
cluster mean-field concepts, which are able to address the real-time dynamics
of strongly correlated lattice fermion models beyond the more simple extreme
limits of one and infinite lattice dimension.  

With the present paper we could
give a proof of principle that a nonequilibrium conserving cluster-perturbation
theory is possible and can be evaluated numerically in practice.  An highly
attractive feature of this approach is the linear scaling with the propagation
time, while the exponential scaling with the cluster size is the typical
bottleneck of any cluster mean-field theory. 

The mapping of the original nonequilibrium CPT onto an effective auxiliary
problem specified by a noninteracting Hamiltonian with additional virtual
(``bath'') degrees of freedom is crucial for the practical implementation of
the approach.  One should note that the number of virtual sites is related to
the number of one-particle excitations and thus grows exponentially with the
original cluster size.  Hence, any practical calculation is limited to a few
(say, at most 10) cluster sites only.  This implies that a systematic
finite-size scaling analysis will be problematic if long-range correlations
dominate the essential physical properties -- this is the above-mentioned
drawback that is shared with any available cluster-mean-field theory.  We
therefore expect that the field of applications of conserving CPT is limited to
problems with possibly strong but short-ranged correlations.

Due to its formulation in terms of Green's functions with time arguments on the
Keldysh contour, the CPT has an inherently causal structure.  With the present
paper we could in particular demonstrate how to exploit this causality for an
efficient time-stepping algorithm where updates of the parameter
renormalization can be limited to the respective last time slice during time
propagation.  The essential problem that had to be solved here consists in
controlling the order (in the sense of a Taylor series) at which the parameter
renormalization on a single time slice enters other physical quantities, such
as the basic time-evolution operator, Green's functions, etc.  This has allowed
us to set up a highly accurate numerical algorithm where conservation laws are
respected with machine precision.

For convenience, first numerical results have been generated for interaction
quenches of the two-dimensional Hubbard model on a square lattice at
half-filling, where particle-number and spin conservation are respected
trivially.  Energy conservation has been enforced by time-dependent
renormalization of the intra-cluster hopping in the $2\times 1$ reference
cluster.  It is worth pointing out that even with this simple approximation
(small cluster) the impact of the self-consistency condition is substantial.
Comparing the conserving against plain CPT, there is a qualitative change of
the time-dependence of the doublon density which is plausible and improves the
theory: Artificial oscillations due to the finite cluster size are almost
completely suppressed, and an ultrafast relaxation to a (prethermal) state with
nearly constant doublon density is predicted as might be expected from previous
computations for one- and infinite-dimensional lattices.

Let us emphasize once more that the purpose of the present paper has been to formally develop the very idea of a constrained CPT and to provide a proof of principle for its practicability. 
There are a couple of future tasks that suggest themselves immediately but are beyond the scope of the present paper:
First of all, a more systematic study of the dependence on the cluster size and shape is needed. 
Note that this also includes the necessity to take into account more than a single optimization parameter as there are four local constraints to be satisfied in the present formulation of the theory [see Eq.\ (\ref{eq:selfcon_cpt})], corresponding to the conservation of spin and particle density as well as two constraints for the doublon density (implying energy conservation).
Hence, for a cluster consisting of $L_{c}$ sites, at most $4L_{c}$ parameters are needed. 
In addition, both the number of constraints and the optimization parameters depend on the spatial symmetries and other symmetries, e.g., particle-hole symmetry, of the original and the reference system.
If necessary, more degrees of freedom and correspondingly more parameters can be generated by coupling uncorrelated ``bath'' sites to the physical sites in the reference system in the spirit of (cluster) dynamical mean-field theory. 
Systematic studies addressing the mentioned issues are necessary before a systematic and quantitative comparison with other approaches or with experiments is meaningful.

Interestingly, the conditional equations for the renormalization parameters feature singular points of non-invertibility.  Technically, this currently restricts our investigations to quenches with small final interaction and short propagation times.  
It is not clear at the moment whether or not a physical meaning can be attributed to those singular points; also this requires further systematic studies.  
According to our present experience, it is well conceivable that, with a suitable regularization scheme, time propagation
through a singularity of the Jacobian is possible and has no apparent impact on the time dependence of physical observables.  
Developing such a regularization scheme is the next task for future studies and the most important issue to make the
conserving CPT a powerful numerical tool to address, e.g., real-time magnetization dynamics, even of inhomogeneous models and on long time scales.

\begin{acknowledgments}
We thank Roman Rausch for providing an exact-diagonalization solver for the
Hubbard model and Felix Hofmann for valuable discussions.  This work has been
supported by the Deutsche Forschungsgemeinschaft through the excellence cluster
``The Hamburg Centre for Ultrafast Imaging - Structure, Dynamics and Control of
Matter at the Atomic Scale'' and through the Sonderforschungsbereich 925
(project B5).  Numerical calculations were performed on the PHYSnet computer
cluster at the University of Hamburg.
\end{acknowledgments}

\appendix

\section{Local constraint on the doublon density}
\label{ap:con_double_occ}

Within this subsection we will use the shorthand notation $G^{(2l)} \equiv
G^{(2l)}_{T,U}$ and $G^{(2r)} \equiv G^{(2r)}_{T,U}$. To prove the local 
constraint on the doublon density, we consider
\begin{align}
  -i\partial_t d_i(t) 
  &=
  \est{[\hat{H}_{T,U}(t),
         \hat{n}_{i\uparrow}(t)\hat{n}_{i\downarrow}(t)]}_{H_{T,U}} \nonumber\\
  &= 
  \sum_{\sigma}
  \est{
    \left[
      \sum_{jk}T_{jk\sigma}(t)\cd{j\sigma}(t)\cc{k\sigma}(t),
      \hat n_{i\sigma}(t) 
    \right]\hat n_{i\bar\sigma}(t)}_{H_{T,U}},
\end{align}
where we used that the double occupation operator commutes with the interaction
term of the Hamiltonian $H_{T,U}$. Using further that
\begin{align}
  \sum_{jk}T_{jk\sigma}(t)
  &\left[
    \cd{j\sigma}(t)\cc{k\sigma}(t),
    \hat n_{i\sigma}(t) 
  \right]\\
  &=
  \sum_{jk}T_{jk\sigma}(t)
  \left(
    \delta_{ki}\cd{j\sigma}(t)\cc{i\sigma}(t)
    -
    \delta_{ji}\cd{i\sigma}(t)\cc{k\sigma}(t)
  \right),\nonumber
\end{align}
we find the final form by comparing with Eqs.\ \eqref{eq:g2l_exact} and
\eqref{eq:g2r_exact} and using the relation $d_i(t) =
-G^{(2l)}_{ii\sigma}(t,t^+) = -G^{(2r)}_{ii\sigma'}(t,t^+)$. This implies
\begin{align}
  \label{eq:double_derived}
  -2i\partial_t d_i(t) 
  &=
  i\partial_t
  \left[
    G^{(2l)}_{ii\sigma}(t,t^+)
    +
    G^{(2r)}_{ii\sigma}(t,t^+)
  \right]\\
  &=
  2\sum_{j\sigma}
  \left[
    T_{ij\sigma}(t)G^{(2r)}_{ji\sigma}(t,t^+)
    -
    G^{(2l)}_{ij\sigma}(t,t^+)T_{ji\sigma}(t)
  \right],\nonumber
\end{align}
which completes our derivation of Eq.\ \eqref{eq:con_energy}.

To prove that Eq. \eqref{eq:double_derived} indeed ensures energy conservation,
let us consider a time-independent Hamiltonian with $T_{ij\sigma}(t) =
T_{ij\sigma}$ and $U(t) = U$.  We consider the time-derivative of the kinetic
energy first. Since the kinetic part of the Hamiltonian trivially commutes with
itself, one obtains
\begin{align}
  i\partial_t E_\text{kin}(t)
  &= 
  \est{
    \sum_{ij\sigma} T_{ij\sigma}\cd{i\sigma}\cc{j\sigma},
    U\sum_l n_{l\uparrow}n_{l\downarrow}
  }\\
  &=
  U
  \sum_{ij\sigma}
  \left[
    T_{ij\sigma}G^{(2r)}_{ji\sigma}(t,t^+)
    -
    G^{(2l)}_{ij\sigma}(t,t^+)T_{ji\sigma}
  \right],\nonumber
\end{align}
This term cancels with $i\partial_t E_\text{int}(t) = U \sum_{i}
i\partial_t d_i(t)$ assuming Eq. \eqref{eq:double_derived} holds thus proving
energy conservation for a time-independent Hamiltonian.

\section{Mathematical structure of the effective Hamiltonian}
\label{ap:math_eff}

In Eq.\ \eqref{eq:heff} we have stated the effective one-particle Hamiltonian
$H^\mathrm{eff}(t) \equiv H^\mathrm{eff}_{T,U}(t)$ for an interacting lattice
system. Its corresponding matrix elements are given by $h(t) \equiv
h_{T,U}(t)$, cf.\  Eq.\ \eqref{eq:h_TU}. In Ref.\ \onlinecite{GP15} it is shown
that $h$ can be constructed from the Lehmann representation of the
one-particle Green's function $G\equiv G_{T,U}$. It is given by
\begin{equation}
  h_{xy\sigma}(t)
  = 
  \sum_{m,n} [i\partial_ t O_{x\sigma (m,n)}(t)] O^*_{y\sigma (m,n)}(t).
\end{equation}
The matrix $O(t)$ can explicitly be stated within the physical sector only
(i.e., $(x,y) = (i,j)$). There it takes the form
\begin{equation}
  \label{eq:defO}
	O_{i\sigma(m,n)}(t) 
  =
  \sqrt{\frac{e^{-\beta E_m}+e^{-\beta E_n}}{Z}} 
  \lwf{m}\cc{i\sigma}(t)\rwf{n},
\end{equation}
where $\rwf{m}$ denotes the $m$-th eigenstate with corresponding eigenenergy
$E_m$ of the initial Hamiltonian, i.e., $H_{T, U}(0)\rwf{m} =
E_m \rwf{m}$. $Z$ denotes the grand-canonical partition function, $\beta$ the
inverse temperature of the equilibrium initial state. The remaining matrix
elements $O_{s\sigma (m,n)}(t)$ are defined uniquely, up to rotations in
invariant subspaces, by requiring $O(t)$ to be a unitary transform, see Ref.\ 
\onlinecite{GP15} for details. Eq.\ \eqref{eq:h=Ueta} is now easily derived
from
\begin{align}
  \label{ap:dtO}
  i\partial_t O_{i\sigma(m,n)}(t)
  &= 
  \sum_{j}
  \left(
    T_{ij\sigma}(t)-\delta_{ij}\mu
  \right)
  O_{j\sigma(m,n)}(t)\\
  &\phantom{=}\,\,\,+
  U(t) R_{i\sigma(m,n)}(t),\nonumber
\end{align}
where
\begin{equation}
  R_{i\sigma(m,n)}(t)
  =
  \sqrt{\frac{e^{-\beta E_m}+e^{-\beta E_n}}{Z}} 
  \lwf{m}\hat{n}_{i\bar\sigma}(t)\cc{i\sigma}(t)\rwf{n}.
\end{equation}
With the definition
\begin{equation}
  \label{eq:ldef}
  \eta_{ix \sigma}(t) = \sum_{mn} R_{i\sigma (m,n)}(t) O_{x\sigma(m,n)}^*(t)
\end{equation}
we arrive at Eq.\ \eqref{eq:h=Ueta}. It is instructive to note that the
elements $\eta_{ij\sigma}(t)$ in the physical sector are easily evaluated as
\begin{align}
  &\eta_{ij\sigma}(t) 
  = 
  \est{
    \hat{n}_{i\bar\sigma}(t)
    \left\{
      \cc{i\sigma}(t),
      \cd{j\sigma}(t)
    \right\}
  }_{H_{T, U}}
  =
  \delta_{ij}
  \est{
    \hat{n}_{i\bar\sigma} (t)
  }_{H_{T, U}},
\end{align}
where $\{ A, B \} = AB + BA$ denotes the anti-commutator.  From a numerical
point of view we note, that $\eta(t)$ as well as its derivatives with respect
to time of arbitrary order can be easily evaluated using the
exact-diagonalization-based scheme presented in Ref.\ \onlinecite{GP15}.

Regarding our discussion of the initial state in Sec.\ \ref{sec:initial_state}
it is now easy to proof that $\mathrm{Im} \{ \eta_{ix\sigma}(0) \} = 0$. At
time $t=0$ the elements $R_{i\sigma(m,n)}(t)$ can be chosen as real, since the
Hamiltonian $H_{T,U}(0)$ is symmetric and therefore has real eigenvectors
$\rwf{m}$. Similarly, the physical sector $O_{i\sigma(m,n)}(t)$ is real in this
case and completion to a unitary matrix $O(t)$ allows for an orthogonal, i.e.,
real $O(t)$. It then follows that $\mathrm{Im} \{ \eta_{ix\sigma}(0) \} = 0$
from Eq.\ \eqref{eq:ldef}. We note that any non-real choice $\tilde O(t)$ can
be brought into the  form $\tilde O(0) = O(0)S$, where the matrix $S$ contains
the phase factors and possibly rotations in invariant subspaces. However, from
Eq.  \eqref{ap:dtO} it follows that $\tilde R(0) = R(0)S$ and therefore the $S$
matrix cancels. The same proof holds for the reference system, i.e.,
$\mathrm{Im}\{\eta'[\lambda]_{ix\sigma}(0)\} = 0$, assuming $\lambda$ is real.

\section{Calculating the time-local variation of $\eta$}
\label{ap:d_eta}

Let $h' = h_{T' - \lambda, U}$ denote the matrix elements of the effective
Hamiltonian of the reference system $H^\mathrm{eff}_{T' - \lambda, U}$, cf.
Eqs.  \eqref{eq:heff} and \eqref{eq:h_TU}. Through Eq.\ \eqref{eq:h=Ueta}, or
equivalently Eq.\  \eqref{eq:ldef}, a corresponding matrix $\eta' \equiv
\eta_{T' - \lambda, U}$ is defined. 
We are interested in how it transforms
under time-local variations.  Since $\eta'$ is an integrated quantity in
$\lambda$, we first calculate its derivative with respect to time.  
Eq.\ \eqref{eq:ldef} implies
\begin{align}
  i\partial_t 
  \eta'_{ix\sigma}(t)
  &=
  -\sum_{y}\eta'_{iy\sigma}(t)h'_{yx\sigma}(t)\\
  &\phantom{=}\,\,\,
  +
  \sum_{mn}
  \left[
    i\partial_t R'_{i\sigma(m,n)}(t)
  \right]
  [O']^\dagger_{(m,n)x\sigma}(t),\nonumber
\end{align}
where $R' \equiv R_{T' - \lambda, U}$ and $O' \equiv O_{T' - \lambda, U}$.
With $H' = H_{T' - \lambda, U}$, the time-local variation of $i\partial_t
R'_{i\sigma (m,n)}(t)$ is given by
\begin{align}
  \deltaloc
  &\left[
    i\partial_t R'_{i\sigma(m,n)}(t)
  \right]
  =
  z^{(m,n)}\lwf{m}\deltaloc
  \left[
    \hat{n}_{i\bar{\sigma}}(t)
    \cc{i\sigma}(t),\hat{H}'(t)
  \right]\rwf{n}\nonumber\\
  &=
  -z^{(m,n)}
  \sum_{j}
  \lwf{m}
  \delta\lambda_{ij\sigma}(t)
  \hat{n}_{i\bar{\sigma}}(t)\cc{j\sigma}(t)\nonumber\\
  &\phantom{=}\quad\quad
  +
  \delta\lambda_{ij\bar\sigma}(t)
  \left[
    \cd{i\bar\sigma}(t)\cc{j\bar\sigma}(t)
    -
    \cd{j\bar\sigma}(t)\cc{i\bar\sigma}(t)
  \right]\cc{i\sigma}(t)
  \rwf{n},
\end{align}
where we further introduced $z^{(m,n)} = \sqrt{(e^{-\beta E_m} + e^{-\beta
E_n})/Z}$ and exploited $\delta \lambda_{ij\sigma} = \delta\lambda_{ji\sigma}$.
The time-local variation of $O'(t)$, on the other hand, vanishes since it is an
integrated quantity in $\lambda$. This follows directly from the definition of
its physical sector, cf.\ Eq.\ \eqref{eq:defO}. We define
\begin{align}
  \gamma_{ix\sigma}^{l\sigma}(t)
  &=
  \sum_{mn}
  z^{(m,n)}
  \lwf{m}
    \hat{n}_{i\bar\sigma}(t)
    \cc{l\sigma}(t)
  \rwf{n}
  [O']^*_{x\sigma (m,n)}(t),\\
  \gamma_{ix\sigma}^{l\bar\sigma}(t)
  &=
  \sum_{mn}
  z^{(m,n)}
  \lwf{m}
    \left[
      \cd{i\bar\sigma}(t)\cc{l\bar\sigma}(t)
    \right.\\
    &\qquad\qquad\qquad
    \left.
    -
      \cd{l\bar\sigma}(t)\cc{i\bar\sigma}(t)
  \right]
  \cc{i\sigma}(t)
  \rwf{n}
  [O']^*_{x\sigma (m,n)}(t),\nonumber
\end{align}
and therewith obtain Eq.\ \eqref{eq:eta_variation}.

\section{High-order time propagation scheme}
\label{sec:high-order}

Finally, we like to set up an efficient numerical scheme to determine
$\lambda^\text{opt}(t)$.  This should be based on a time-propagation algorithm
where the error is of high order in the basic time step $\Delta t$.  Let us
assume that for each time step the Taylor expansion of $\lambda^\text{opt}(t)$
is well defined.  For each time interval and for arbitrary $t \in [t_{n},
t_{n+1}]$ we then have
\begin{equation}
  \lambda^\text{opt}(t) 
  =
  \left\{
    \begin{aligned}
      &\lambda^{(0)}(t) 
      = 
      \sum_{p = 0}^{n_p} \frac{\lambda_{0,p}}{p!} t^p + O(\Delta t^{n_p + 1})
      \quad\text{if} \;  t\in[0, t_1[,\\
          &\lambda^{(1)}(t) 
      = 
      \sum_{p = 0}^{n_p} \frac{\lambda_{1,p}}{p!} 
      (t - t_1)^p + O(\Delta t^{n_p + 1})\\
      &\,\,\,\,\,\quad\qquad\qquad\qquad\qquad\qquad\qquad
       \text{if} \; t\in[t_1, t_2[,\\
      &\dots
    \end{aligned}
  \right.
\end{equation}
where each $\lambda$-term must be considered as a tuple with components
labelled by the super-index $b$, e.g., $\lambda_{n,p} = ([\lambda_{n,p}]_b)$,
where $n$ refers to the $n$-th time interval, and where $p$ runs from $p=0$ up
to the maximum order of the polynomial $n_p$.  During the time propagation, the
polynomial approximation must be updated after each time step.  This is done by
fixing the coefficients at each interfacing time $t_n$ such that
$\Gamma[\lambda^\text{opt}](t_n) = 0$. For times $t \neq t_n$ we then have
$\Gamma[\lambda^\text{opt}](t) = O(\Delta t^{n_p + 1})$.  Writing $J(t) \equiv
J[\lambda^\text{opt}](t)$ and $\xi_\Gamma(t)\equiv
\xi_\Gamma[\lambda^\text{opt}](t)$ for short and applying the product rule to
$J(t) \lambda^\text{opt}(t) = \xi_\Gamma(t)$, the self-consistency condition
\eqref{eq:lopt} is readily rewritten in terms of the Taylor coefficients:
\begin{equation}
  \label{eq:lHn}
  \lambda_{n,p} 
  = 
  J^{-1}(t_n)
  \left(
    \sum_{r=0}^{p-1} \binom{p}{r} [\partial_t^{p-r} J(t)]_{t=t_n} 
    \lambda_{n,r}
    - 
    [\partial_t^p \xi_{\Gamma}(t)]_{t =t_n}
  \right).
\end{equation}

Suppose that $\lambda^{(q)}(t)$ is known for all $q < n$, i.e., suppose that
the propagation has been completed over the interval $[0, t_n[$.  The next step
is to update the coefficients.  At this point we can exploit that $J(t)$ and
$\xi_\Gamma(t)$ scale like integrated quantities in $\lambda$ under time-local
variations which implies $\deltaloc J(t) = 0$ and $\deltaloc \xi_\Gamma(t) =
0$. Hence, at $t=t_n$, both are independent of $\lambda^\text{opt}(t_n)$. We
define
\begin{equation}
  \tilde \lambda(t)
  =
  \left\{
    \begin{aligned}
      &\lambda^\text{opt}(t)\quad \text{if} \quad t < t_n,\\
      &0 \quad \text{else}.
    \end{aligned}
  \right.
\end{equation}
Then, 
\begin{equation}
  \xi_\Gamma(t_n) = \Gamma[\tilde \lambda](t_n),
\end{equation}
and we are now able to solve Eq.\ \eqref{eq:lHn} for $\lambda_{n,0}$. 
The first derivatives $\partial_t J(t)\bigr|_{t=t_n}$ and $\partial_t
\xi_\Gamma(t)\bigr|_{t=t_n}$ explicitly depend on $\lambda^\text{opt}(t_n) =
\lambda_{n,0}$, which is now known, but are integrated quantities in the first
derivative $\partial_t \lambda^\text{opt}(t)$, i.e., they are independent of
$\partial_t \lambda^\text{opt}(t)\bigr|_{t=t_n} = \lambda_{n,1}$.  Therefore,
the same idea can be applied and in fact be repeated again and again until
finally $\lambda^{(n)}(t)$ is known up to the desired order.  We emphasize that
the presented algorithm gives a fully converged $\lambda^\text{opt}(t) =
\lambda^{(n)}(t) + O(\Delta t^{n_p + 1})$ for $t \in [t_n, t_{n+1}[$ within a
single iteration.


\begin{thebibliography}{51}
\expandafter\ifx\csname natexlab\endcsname\relax\def\natexlab#1{#1}\fi
\expandafter\ifx\csname bibnamefont\endcsname\relax
  \def\bibnamefont#1{#1}\fi
\expandafter\ifx\csname bibfnamefont\endcsname\relax
  \def\bibfnamefont#1{#1}\fi
\expandafter\ifx\csname citenamefont\endcsname\relax
  \def\citenamefont#1{#1}\fi
\expandafter\ifx\csname url\endcsname\relax
  \def\url#1{\texttt{#1}}\fi
\expandafter\ifx\csname urlprefix\endcsname\relax\def\urlprefix{URL }\fi
\providecommand{\bibinfo}[2]{#2}
\providecommand{\eprint}[2][]{\url{#2}}

\bibitem[{\citenamefont{Polkovnikov et~al.}(2011)\citenamefont{Polkovnikov,
  Sengupta, Silva, and Vengalattore}}]{PSSV11}
\bibinfo{author}{\bibfnamefont{A.}~\bibnamefont{Polkovnikov}},
  \bibinfo{author}{\bibfnamefont{K.}~\bibnamefont{Sengupta}},
  \bibinfo{author}{\bibfnamefont{A.}~\bibnamefont{Silva}}, \bibnamefont{and}
  \bibinfo{author}{\bibfnamefont{M.}~\bibnamefont{Vengalattore}},
  \bibinfo{journal}{Rev. Mod. Phys.} \textbf{\bibinfo{volume}{83}},
  \bibinfo{pages}{863} (\bibinfo{year}{2011}).

\bibitem[{\citenamefont{Aoki et~al.}(2014)\citenamefont{Aoki, Tsuji, Eckstein,
  Kollar, Oka, and Werner}}]{ATE+14}
\bibinfo{author}{\bibfnamefont{H.}~\bibnamefont{Aoki}},
  \bibinfo{author}{\bibfnamefont{N.}~\bibnamefont{Tsuji}},
  \bibinfo{author}{\bibfnamefont{M.}~\bibnamefont{Eckstein}},
  \bibinfo{author}{\bibfnamefont{M.}~\bibnamefont{Kollar}},
  \bibinfo{author}{\bibfnamefont{T.}~\bibnamefont{Oka}}, \bibnamefont{and}
  \bibinfo{author}{\bibfnamefont{P.}~\bibnamefont{Werner}},
  \bibinfo{journal}{Rev. Mod. Phys.} \textbf{\bibinfo{volume}{86}},
  \bibinfo{pages}{779} (\bibinfo{year}{2014}).

\bibitem[{\citenamefont{Hochbruck and Lubich}(1997)}]{HL97}
\bibinfo{author}{\bibfnamefont{M.}~\bibnamefont{Hochbruck}} \bibnamefont{and}
  \bibinfo{author}{\bibfnamefont{C.}~\bibnamefont{Lubich}},
  \bibinfo{journal}{SIAM J. Numerical Anal.} \textbf{\bibinfo{volume}{34}},
  \bibinfo{pages}{1911} (\bibinfo{year}{1997}).

\bibitem[{\citenamefont{Blankenbecler et~al.}(1981)\citenamefont{Blankenbecler,
  Scalapino, and Sugar}}]{BSS81}
\bibinfo{author}{\bibfnamefont{R.}~\bibnamefont{Blankenbecler}},
  \bibinfo{author}{\bibfnamefont{D.}~\bibnamefont{Scalapino}},
  \bibnamefont{and} \bibinfo{author}{\bibfnamefont{R.~L.} \bibnamefont{Sugar}},
  \bibinfo{journal}{Phys. Rev. D} \textbf{\bibinfo{volume}{8}},
  \bibinfo{pages}{2278} (\bibinfo{year}{1981}).

\bibitem[{\citenamefont{White et~al.}(1989)\citenamefont{White, Scalapino,
  Sugar, and Bickers}}]{WSSB89}
\bibinfo{author}{\bibfnamefont{S.~R.} \bibnamefont{White}},
  \bibinfo{author}{\bibfnamefont{D.~J.} \bibnamefont{Scalapino}},
  \bibinfo{author}{\bibfnamefont{R.~L.} \bibnamefont{Sugar}}, \bibnamefont{and}
  \bibinfo{author}{\bibfnamefont{N.~E.} \bibnamefont{Bickers}},
  \bibinfo{journal}{Phys. Rev. Lett.} \textbf{\bibinfo{volume}{63}},
  \bibinfo{pages}{1523} (\bibinfo{year}{1989}).

\bibitem[{\citenamefont{Assaad and Evertz}(2008)}]{AE08}
\bibinfo{author}{\bibfnamefont{F.~F.} \bibnamefont{Assaad}} \bibnamefont{and}
  \bibinfo{author}{\bibfnamefont{H.~G.} \bibnamefont{Evertz}}, in:
  Computational Many-Particle Physics, Vol. 739 of Lecture Notes in Physics,
  edited by H. Fehske, R. Schneider, and A. Wei\ss{}e, pp.\ 277
  (\bibinfo{publisher}{Springer}, \bibinfo{address}{Berlin},
  \bibinfo{year}{2008}).

\bibitem[{\citenamefont{M\"uhlbacher and Rabani}(2008)}]{MR08}
\bibinfo{author}{\bibfnamefont{L.}~\bibnamefont{M\"uhlbacher}}
  \bibnamefont{and} \bibinfo{author}{\bibfnamefont{E.}~\bibnamefont{Rabani}},
  \bibinfo{journal}{Phys. Rev. Lett.} \textbf{\bibinfo{volume}{100}},
  \bibinfo{pages}{176403} (\bibinfo{year}{2008}).

\bibitem[{\citenamefont{P.~Werner}(2009)}]{WOM09}
P.~Werner, T.~Oka, and A.~J.~Millis, Phys. Rev. B
  \textbf{\bibinfo{volume}{79}}, \bibinfo{pages}{035320}
  (\bibinfo{year}{2009}).

\bibitem[{\citenamefont{Schir\'o and Fabrizio}(2009)}]{SF09}
\bibinfo{author}{\bibfnamefont{M.}~\bibnamefont{Schir\'o}} \bibnamefont{and}
  \bibinfo{author}{\bibfnamefont{M.}~\bibnamefont{Fabrizio}},
  \bibinfo{journal}{Phys. Rev. B} \textbf{\bibinfo{volume}{79}},
  \bibinfo{pages}{153302} (\bibinfo{year}{2009}).

\bibitem[{\citenamefont{Cohen et~al.}(2015)\citenamefont{Cohen, Gull, Reichman,
  and Millis}}]{CGRM15}
\bibinfo{author}{\bibfnamefont{G.}~\bibnamefont{Cohen}},
  \bibinfo{author}{\bibfnamefont{E.}~\bibnamefont{Gull}},
  \bibinfo{author}{\bibfnamefont{D.~R.} \bibnamefont{Reichman}},
  \bibnamefont{and} \bibinfo{author}{\bibfnamefont{A.~J.}
  \bibnamefont{Millis}}, \bibinfo{journal}{Phys. Rev. Lett.}
  \textbf{\bibinfo{volume}{115}}, \bibinfo{pages}{266802}
  (\bibinfo{year}{2015}).

\bibitem[{\citenamefont{Anders and Schiller}(2005)}]{AS05}
\bibinfo{author}{\bibfnamefont{F. B.}~\bibnamefont{Anders}} \bibnamefont{and}
  \bibinfo{author}{\bibfnamefont{A.}~\bibnamefont{Schiller}},
  \bibinfo{journal}{Phys. Rev. Lett.} \textbf{\bibinfo{volume}{95}},
  \bibinfo{pages}{196801} (\bibinfo{year}{2005}).

\bibitem[{\citenamefont{White and Feiguin}(2004)}]{WF04}
\bibinfo{author}{\bibfnamefont{S.~R.} \bibnamefont{White}} \bibnamefont{and}
  \bibinfo{author}{\bibfnamefont{A.~E.} \bibnamefont{Feiguin}},
  \bibinfo{journal}{Phys. Rev. Lett.} \textbf{\bibinfo{volume}{93}},
  \bibinfo{pages}{076401} (\bibinfo{year}{2004}).

\bibitem[{\citenamefont{Vidal}(2004)}]{Vid04}
\bibinfo{author}{\bibfnamefont{G.}~\bibnamefont{Vidal}},
  \bibinfo{journal}{Phys. Rev. Lett.} \textbf{\bibinfo{volume}{93}},
  \bibinfo{pages}{40502} (\bibinfo{year}{2004}).

\bibitem[{\citenamefont{Haegeman et~al.}(2016)\citenamefont{Haegeman, Lubich,
  Oseledets, Vandereycken, and Verstraete}}]{HLO+16}
\bibinfo{author}{\bibfnamefont{J.}~\bibnamefont{Haegeman}},
  \bibinfo{author}{\bibfnamefont{C.}~\bibnamefont{Lubich}},
  \bibinfo{author}{\bibfnamefont{I.}~\bibnamefont{Oseledets}},
  \bibinfo{author}{\bibfnamefont{B.}~\bibnamefont{Vandereycken}},
  \bibnamefont{and}
  \bibinfo{author}{\bibfnamefont{F.}~\bibnamefont{Verstraete}},
  \bibinfo{journal}{Phys. Rev. B} \textbf{\bibinfo{volume}{94}},
  \bibinfo{pages}{165116} (\bibinfo{year}{2016}).

\bibitem[{\citenamefont{Sandri et~al.}(2012)\citenamefont{Sandri, Schir\'o, and
  Fabrizio}}]{SSF12}
\bibinfo{author}{\bibfnamefont{M.}~\bibnamefont{Sandri}},
  \bibinfo{author}{\bibfnamefont{M.}~\bibnamefont{Schir\'o}}, \bibnamefont{and}
  \bibinfo{author}{\bibfnamefont{M.}~\bibnamefont{Fabrizio}},
  \bibinfo{journal}{Phys. Rev. B} \textbf{\bibinfo{volume}{86}},
  \bibinfo{pages}{075122} (\bibinfo{year}{2012}).

\bibitem[{\citenamefont{Ido et~al.}(2015)\citenamefont{Ido, Ohgoe, and
  Imada}}]{IOI15}
\bibinfo{author}{\bibfnamefont{K.}~\bibnamefont{Ido}},
  \bibinfo{author}{\bibfnamefont{T.}~\bibnamefont{Ohgoe}}, \bibnamefont{and}
  \bibinfo{author}{\bibfnamefont{M.}~\bibnamefont{Imada}},
  \bibinfo{journal}{Phys. Rev. B} \textbf{\bibinfo{volume}{92}},
  \bibinfo{pages}{245106} (\bibinfo{year}{2015}).

\bibitem[{\citenamefont{Carleo et~al.}(2012)\citenamefont{Carleo, Becca,
  Schir\'o, and Fabrizio}}]{CBSF12}
\bibinfo{author}{\bibfnamefont{G.}~\bibnamefont{Carleo}},
  \bibinfo{author}{\bibfnamefont{F.}~\bibnamefont{Becca}},
  \bibinfo{author}{\bibfnamefont{M.}~\bibnamefont{Schir\'o}}, \bibnamefont{and}
  \bibinfo{author}{\bibfnamefont{M.}~\bibnamefont{Fabrizio}},
  \bibinfo{journal}{Sci. Rep.} \textbf{\bibinfo{volume}{2}},
  \bibinfo{pages}{243} (\bibinfo{year}{2012}).

\bibitem[{\citenamefont{Baym and Kadanoff}(1961)}]{BK61}
\bibinfo{author}{\bibfnamefont{G.}~\bibnamefont{Baym}} \bibnamefont{and}
  \bibinfo{author}{\bibfnamefont{L.~P.} \bibnamefont{Kadanoff}},
  \bibinfo{journal}{Phys. Rev.} \textbf{\bibinfo{volume}{124}},
  \bibinfo{pages}{287} (\bibinfo{year}{1961}).

\bibitem[{\citenamefont{Baym}(1962)}]{Bay62}
\bibinfo{author}{\bibfnamefont{G.}~\bibnamefont{Baym}}, \bibinfo{journal}{Phys.
  Rev.} \textbf{\bibinfo{volume}{127}}, \bibinfo{pages}{1391}
  (\bibinfo{year}{1962}).

\bibitem[{\citenamefont{Luttinger and Ward}(1960)}]{LW60}
\bibinfo{author}{\bibfnamefont{J.~M.} \bibnamefont{Luttinger}}
  \bibnamefont{and} \bibinfo{author}{\bibfnamefont{J.~C.} \bibnamefont{Ward}},
  \bibinfo{journal}{Phys. Rev.} \textbf{\bibinfo{volume}{118}},
  \bibinfo{pages}{1417} (\bibinfo{year}{1960}).

\bibitem[{\citenamefont{Bickers et~al.}(1989)\citenamefont{Bickers, Scalapino,
  and White}}]{BSW89}
\bibinfo{author}{\bibfnamefont{N.~E.} \bibnamefont{Bickers}},
  \bibinfo{author}{\bibfnamefont{D.~J.} \bibnamefont{Scalapino}},
  \bibnamefont{and} \bibinfo{author}{\bibfnamefont{S.~R.} \bibnamefont{White}},
  \bibinfo{journal}{Phys. Rev. Lett.} \textbf{\bibinfo{volume}{62}},
  \bibinfo{pages}{961} (\bibinfo{year}{1989}).

\bibitem[{\citenamefont{Joura et~al.}(2015)\citenamefont{Joura, Freericks, and
  Lichtenstein}}]{Joura:15}
\bibinfo{author}{\bibfnamefont{A.~V.} \bibnamefont{Joura}},
  \bibinfo{author}{\bibfnamefont{J.~K.} \bibnamefont{Freericks}},
  \bibnamefont{and} \bibinfo{author}{\bibfnamefont{A.~I.}
  \bibnamefont{Lichtenstein}}, \bibinfo{journal}{Phys. Rev. B}
  \textbf{\bibinfo{volume}{91}}, \bibinfo{pages}{245153}
  (\bibinfo{year}{2015}).

\bibitem[{\citenamefont{Hofmann et~al.}(2013)\citenamefont{Hofmann, Eckstein,
  Arrigoni, and Potthoff}}]{HEAP13}
\bibinfo{author}{\bibfnamefont{F.}~\bibnamefont{Hofmann}},
  \bibinfo{author}{\bibfnamefont{M.}~\bibnamefont{Eckstein}},
  \bibinfo{author}{\bibfnamefont{E.}~\bibnamefont{Arrigoni}}, \bibnamefont{and}
  \bibinfo{author}{\bibfnamefont{M.}~\bibnamefont{Potthoff}},
  \bibinfo{journal}{Phys. Rev. B} \textbf{\bibinfo{volume}{88}},
  \bibinfo{pages}{165124} (\bibinfo{year}{2013}).

\bibitem[{\citenamefont{Potthoff}(2003{\natexlab{a}})}]{Pot03a}
\bibinfo{author}{\bibfnamefont{M.}~\bibnamefont{Potthoff}},
  \bibinfo{journal}{Euro. Phys. J. B} \textbf{\bibinfo{volume}{32}},
  \bibinfo{pages}{429} (\bibinfo{year}{2003}{\natexlab{a}}).

\bibitem[{\citenamefont{Potthoff}(2012)}]{Pot12}
\bibinfo{author}{\bibfnamefont{M.}~\bibnamefont{Potthoff}},
  \emph{\bibinfo{title}{{\rm In:} Strongly Correlated Systems: Theoretical
  Methods}}, Ed. by A. Avella and F. Mancini, Springer Series in Solid-State
  Sciences, Vol. 171, p. 303 (\bibinfo{publisher}{Springer},
  \bibinfo{address}{Berlin}, \bibinfo{year}{2012}).

\bibitem[{\citenamefont{Potthoff et~al.}(2003)\citenamefont{Potthoff, Aichhorn,
  and Dahnken}}]{PAD03}
\bibinfo{author}{\bibfnamefont{M.}~\bibnamefont{Potthoff}},
  \bibinfo{author}{\bibfnamefont{M.}~\bibnamefont{Aichhorn}}, \bibnamefont{and}
  \bibinfo{author}{\bibfnamefont{C.}~\bibnamefont{Dahnken}},
  \bibinfo{journal}{Phys. Rev. Lett.} \textbf{\bibinfo{volume}{91}},
  \bibinfo{pages}{206402} (\bibinfo{year}{2003}).

\bibitem[{\citenamefont{Dahnken et~al.}(2004)\citenamefont{Dahnken, Aichhorn,
  Hanke, Arrigoni, and Potthoff}}]{DAH+04}
\bibinfo{author}{\bibfnamefont{C.}~\bibnamefont{Dahnken}},
  \bibinfo{author}{\bibfnamefont{M.}~\bibnamefont{Aichhorn}},
  \bibinfo{author}{\bibfnamefont{W.}~\bibnamefont{Hanke}},
  \bibinfo{author}{\bibfnamefont{E.}~\bibnamefont{Arrigoni}}, \bibnamefont{and}
  \bibinfo{author}{\bibfnamefont{M.}~\bibnamefont{Potthoff}},
  \bibinfo{journal}{Phys. Rev. B} \textbf{\bibinfo{volume}{70}},
  \bibinfo{pages}{245110} (\bibinfo{year}{2004}).

\bibitem[{\citenamefont{Potthoff}(2003{\natexlab{b}})}]{Pot03b}
\bibinfo{author}{\bibfnamefont{M.}~\bibnamefont{Potthoff}},
  \bibinfo{journal}{Euro. Phys. J. B} \textbf{\bibinfo{volume}{36}},
  \bibinfo{pages}{335} (\bibinfo{year}{2003}{\natexlab{b}}).

\bibitem[{\citenamefont{Eckstein et~al.}(2009)\citenamefont{Eckstein, Kollar,
  and Werner}}]{EKW09}
\bibinfo{author}{\bibfnamefont{M.}~\bibnamefont{Eckstein}},
  \bibinfo{author}{\bibfnamefont{M.}~\bibnamefont{Kollar}}, \bibnamefont{and}
  \bibinfo{author}{\bibfnamefont{P.}~\bibnamefont{Werner}},
  \bibinfo{journal}{Phys. Rev. Lett.} \textbf{\bibinfo{volume}{103}},
  \bibinfo{pages}{056403} (\bibinfo{year}{2009}).

\bibitem[{\citenamefont{Hofmann
  et~al.}(2016{\natexlab{a}})\citenamefont{Hofmann, Eckstein, and
  Potthoff}}]{HEP16a}
\bibinfo{author}{\bibfnamefont{F.}~\bibnamefont{Hofmann}},
  \bibinfo{author}{\bibfnamefont{M.}~\bibnamefont{Eckstein}}, \bibnamefont{and}
  \bibinfo{author}{\bibfnamefont{M.}~\bibnamefont{Potthoff}},
  \bibinfo{journal}{J. Phys.: Conf. Ser.} \textbf{\bibinfo{volume}{696}},
  \bibinfo{pages}{012002} (\bibinfo{year}{2016}{\natexlab{a}}).

\bibitem[{\citenamefont{Hofmann
  et~al.}(2016{\natexlab{b}})\citenamefont{Hofmann, Eckstein, and
  Potthoff}}]{HEP16b}
\bibinfo{author}{\bibfnamefont{F.}~\bibnamefont{Hofmann}},
  \bibinfo{author}{\bibfnamefont{M.}~\bibnamefont{Eckstein}}, \bibnamefont{and}
  \bibinfo{author}{\bibfnamefont{M.}~\bibnamefont{Potthoff}},
  \bibinfo{journal}{Phys. Rev. B} \textbf{\bibinfo{volume}{93}},
  \bibinfo{pages}{235104} (\bibinfo{year}{2016}{\natexlab{b}}).

\bibitem[{\citenamefont{Hofmann and Potthoff}(2016)}]{HP16}
\bibinfo{author}{\bibfnamefont{F.}~\bibnamefont{Hofmann}} \bibnamefont{and}
  \bibinfo{author}{\bibfnamefont{M.}~\bibnamefont{Potthoff}},
  \bibinfo{journal}{Euro. Phys. J. B} \textbf{\bibinfo{volume}{89}},
  \bibinfo{pages}{178} (\bibinfo{year}{2016}).

\bibitem[{\citenamefont{Schmidt and Monien}(2002)}]{SM02}
\bibinfo{author}{\bibfnamefont{P.}~\bibnamefont{Schmidt}} \bibnamefont{and}
  \bibinfo{author}{\bibfnamefont{H.}~\bibnamefont{Monien}},
  cond-mat/0202046.

\bibitem[{\citenamefont{Freericks et~al.}(2006)\citenamefont{Freericks,
  Turkowski, and Zlati\'c}}]{FTZ06}
\bibinfo{author}{\bibfnamefont{J.~K.} \bibnamefont{Freericks}},
  \bibinfo{author}{\bibfnamefont{V.~M.} \bibnamefont{Turkowski}},
  \bibnamefont{and} \bibinfo{author}{\bibfnamefont{V.}~\bibnamefont{Zlati\'c}},
  \bibinfo{journal}{Phys. Rev. Lett.} \textbf{\bibinfo{volume}{97}},
  \bibinfo{pages}{266408} (\bibinfo{year}{2006}).

\bibitem[{\citenamefont{Tsuji et~al.}(2014)\citenamefont{Tsuji, Barmettler,
  Aoki, and Werner}}]{TBAP14}
\bibinfo{author}{\bibfnamefont{N.}~\bibnamefont{Tsuji}},
  \bibinfo{author}{\bibfnamefont{P.}~\bibnamefont{Barmettler}},
  \bibinfo{author}{\bibfnamefont{H.}~\bibnamefont{Aoki}}, \bibnamefont{and}
  \bibinfo{author}{\bibfnamefont{P.}~\bibnamefont{Werner}},
  \bibinfo{journal}{Phys. Rev. B} \textbf{\bibinfo{volume}{90}},
  \bibinfo{pages}{075117} (\bibinfo{year}{2014}).

\bibitem[{\citenamefont{Herrmann et~al.}(2016)\citenamefont{Herrmann, Tsuji,
  Eckstein, and Werner}}]{HTEP16}
\bibinfo{author}{\bibfnamefont{A.~J.} \bibnamefont{Herrmann}},
  \bibinfo{author}{\bibfnamefont{N.}~\bibnamefont{Tsuji}},
  \bibinfo{author}{\bibfnamefont{M.}~\bibnamefont{Eckstein}}, \bibnamefont{and}
  \bibinfo{author}{\bibfnamefont{P.}~\bibnamefont{Werner}},
  \bibinfo{journal}{Phys. Rev. B} \textbf{\bibinfo{volume}{94}},
  \bibinfo{pages}{245114} (\bibinfo{year}{2016}).

\bibitem[{\citenamefont{Balzer and Potthoff}(2011)}]{BP11}
\bibinfo{author}{\bibfnamefont{M.}~\bibnamefont{Balzer}} \bibnamefont{and}
  \bibinfo{author}{\bibfnamefont{M.}~\bibnamefont{Potthoff}},
  \bibinfo{journal}{Phys. Rev. B} \textbf{\bibinfo{volume}{83}},
  \bibinfo{pages}{195132} (\bibinfo{year}{2011}).

\bibitem[{\citenamefont{Jurgenowski and Potthoff}(2013)}]{JP13}
\bibinfo{author}{\bibfnamefont{P.}~\bibnamefont{Jurgenowski}} \bibnamefont{and}
  \bibinfo{author}{\bibfnamefont{M.}~\bibnamefont{Potthoff}},
  \bibinfo{journal}{Phys. Rev. B} \textbf{\bibinfo{volume}{87}},
  \bibinfo{pages}{205118} (\bibinfo{year}{2013}).

\bibitem[{\citenamefont{Gramsch and Potthoff}(2015)}]{GP15}
\bibinfo{author}{\bibfnamefont{C.}~\bibnamefont{Gramsch}} \bibnamefont{and}
  \bibinfo{author}{\bibfnamefont{M.}~\bibnamefont{Potthoff}},
  \bibinfo{journal}{Phys. Rev. B} \textbf{\bibinfo{volume}{92}},
  \bibinfo{pages}{235135} (\bibinfo{year}{2015}).

\bibitem[{\citenamefont{S\'en\'echal et~al.}(2000)\citenamefont{S\'en\'echal,
  P\'erez, and Pioro-Ladri\`ere}}]{SPPL00}
\bibinfo{author}{\bibfnamefont{D.}~\bibnamefont{S\'en\'echal}},
  \bibinfo{author}{\bibfnamefont{D.}~\bibnamefont{P\'erez}}, \bibnamefont{and}
  \bibinfo{author}{\bibfnamefont{M.}~\bibnamefont{Pioro-Ladri\`ere}},
  \bibinfo{journal}{Phys. Rev. Lett.} \textbf{\bibinfo{volume}{84}},
  \bibinfo{pages}{522} (\bibinfo{year}{2000}).

\bibitem[{\citenamefont{Gros and Valenti}(1993)}]{GV93}
\bibinfo{author}{\bibfnamefont{C.}~\bibnamefont{Gros}} \bibnamefont{and}
  \bibinfo{author}{\bibfnamefont{R.}~\bibnamefont{Valenti}},
  \bibinfo{journal}{Phys. Rev. B} \textbf{\bibinfo{volume}{48}},
  \bibinfo{pages}{418} (\bibinfo{year}{1993}).

\bibitem[{\citenamefont{Gramsch et~al.}(2013)\citenamefont{Gramsch, Balzer,
  Eckstein, and Kollar}}]{GBEK13}
\bibinfo{author}{\bibfnamefont{C.}~\bibnamefont{Gramsch}},
  \bibinfo{author}{\bibfnamefont{K.}~\bibnamefont{Balzer}},
  \bibinfo{author}{\bibfnamefont{M.}~\bibnamefont{Eckstein}}, \bibnamefont{and}
  \bibinfo{author}{\bibfnamefont{M.}~\bibnamefont{Kollar}},
  \bibinfo{journal}{Phys. Rev. B} \textbf{\bibinfo{volume}{88}},
  \bibinfo{pages}{235106} (\bibinfo{year}{2013}).

\bibitem[{\citenamefont{Balzer and Eckstein}(2014)}]{BE14}
\bibinfo{author}{\bibfnamefont{K.}~\bibnamefont{Balzer}} \bibnamefont{and}
  \bibinfo{author}{\bibfnamefont{M.}~\bibnamefont{Eckstein}},
  \bibinfo{journal}{Phys. Rev. B} \textbf{\bibinfo{volume}{89}},
  \bibinfo{pages}{035148} (\bibinfo{year}{2014}).

\bibitem[{\citenamefont{Perfetti et~al.}(2006)\citenamefont{Perfetti, Loukakos,
  Lisowski, Bovensiepen, Berger, Biermann, Cornaglia, Georges, and
  Wolf}}]{perf:06}
\bibinfo{author}{\bibfnamefont{L.}~\bibnamefont{Perfetti}},
  \bibinfo{author}{\bibfnamefont{P.~A.} \bibnamefont{Loukakos}},
  \bibinfo{author}{\bibfnamefont{M.}~\bibnamefont{Lisowski}},
  \bibinfo{author}{\bibfnamefont{U.}~\bibnamefont{Bovensiepen}},
  \bibinfo{author}{\bibfnamefont{H.}~\bibnamefont{Berger}},
  \bibinfo{author}{\bibfnamefont{S.}~\bibnamefont{Biermann}},
  \bibinfo{author}{\bibfnamefont{P.~S.} \bibnamefont{Cornaglia}},
  \bibinfo{author}{\bibfnamefont{A.}~\bibnamefont{Georges}}, \bibnamefont{and}
  \bibinfo{author}{\bibfnamefont{M.}~\bibnamefont{Wolf}},
  \bibinfo{journal}{Phys. Rev. Lett.} \textbf{\bibinfo{volume}{97}},
  \bibinfo{pages}{067402} (\bibinfo{year}{2006}).

\bibitem[{\citenamefont{Bloch et~al.}(2008)\citenamefont{Bloch, Dalibard, and
  Zwerger}}]{bloch:08}
\bibinfo{author}{\bibfnamefont{I.}~\bibnamefont{Bloch}},
  \bibinfo{author}{\bibfnamefont{J.}~\bibnamefont{Dalibard}}, \bibnamefont{and}
  \bibinfo{author}{\bibfnamefont{W.}~\bibnamefont{Zwerger}},
  \bibinfo{journal}{Rev. Mod. Phys.} \textbf{\bibinfo{volume}{80}},
  \bibinfo{pages}{885} (\bibinfo{year}{2008}).

\bibitem[{\citenamefont{Keldysh}(1964)}]{kel:64}
\bibinfo{author}{\bibfnamefont{L.~V.} \bibnamefont{Keldysh}},
  \bibinfo{journal}{J. Exptl. Theoret. Phys.} \textbf{\bibinfo{volume}{47}},
  \bibinfo{pages}{1515} (\bibinfo{year}{1964}) [Sov. Phys. JETP 20, 1018 (1965)]. 

\bibitem[{\citenamefont{Rammer}(2007)}]{ram:07}
\bibinfo{author}{\bibfnamefont{J.}~\bibnamefont{Rammer}},
  \emph{\bibinfo{title}{Quantum Field Theory of Non-equilibrium States}}
  (\bibinfo{publisher}{Cambridge University Press},
  \bibinfo{address}{Cambridge, UK}, \bibinfo{year}{2007}).

\bibitem[{\citenamefont{van Leeuwen et~al.}(2006)\citenamefont{van Leeuwen,
  Dahlen, Stefanucci, Almbladh, and von Barth}}]{lee:06}
\bibinfo{author}{\bibfnamefont{R.}~\bibnamefont{van Leeuwen}},
  \bibinfo{author}{\bibfnamefont{N.~E.} \bibnamefont{Dahlen}},
  \bibinfo{author}{\bibfnamefont{G.}~\bibnamefont{Stefanucci}},
  \bibinfo{author}{\bibfnamefont{C.-O.} \bibnamefont{Almbladh}},
  \bibnamefont{and} \bibinfo{author}{\bibfnamefont{U.}~\bibnamefont{von
  Barth}}, \emph{\bibinfo{title}{Introduction to the Keldysh formalism}}, vol.
  \bibinfo{volume}{706} of \emph{\bibinfo{series}{Lecture Notes in Physics}}
  (\bibinfo{publisher}{Springer}, \bibinfo{address}{Heidelberg, Germany},
  \bibinfo{year}{2006}).

\bibitem[{\citenamefont{Rausch and Potthoff}(2016)}]{RP17}
R. Rausch and M. Potthoff, 
Phys. Rev. B {\bf 95}, 045152 (2017)

\bibitem[{\citenamefont{Moeckel and Kehrein}(2008)}]{MK08}
\bibinfo{author}{\bibfnamefont{M.}~\bibnamefont{Moeckel}} \bibnamefont{and}
  \bibinfo{author}{\bibfnamefont{S.}~\bibnamefont{Kehrein}},
  \bibinfo{journal}{Phys. Rev. Lett.} \textbf{\bibinfo{volume}{100}},
  \bibinfo{pages}{175702} (\bibinfo{year}{2008}).

\bibitem[{\citenamefont{Kennes et~al.}(2017)\citenamefont{Kennes, Pommerening,
  Diekmann, Karrasch, and Meden}}]{KPD+17}
\bibinfo{author}{\bibfnamefont{D.~M.} \bibnamefont{Kennes}},
  \bibinfo{author}{\bibfnamefont{J.~C.} \bibnamefont{Pommerening}},
  \bibinfo{author}{\bibfnamefont{J.}~\bibnamefont{Diekmann}},
  \bibinfo{author}{\bibfnamefont{C.}~\bibnamefont{Karrasch}}, \bibnamefont{and}
  \bibinfo{author}{\bibfnamefont{V.}~\bibnamefont{Meden}},
  \bibinfo{journal}{Phys. Rev. B} \textbf{\bibinfo{volume}{95}},
  \bibinfo{pages}{035147} (\bibinfo{year}{2017}).

\end{thebibliography}

\end{document}